\documentclass[prl,twocolumn,altaffilletter,nolongbibliography,numerical,flushbottom,secnumarabic,superscriptaddress,floatfix]{revtex4-2}

\usepackage{graphicx}
\usepackage{braket}
\usepackage{booktabs}  
\usepackage{comment}
\usepackage[normalem]{ulem}
\usepackage[title]{appendix}
\usepackage{amsmath,mathtools,amsthm,amssymb,pifont}
\usepackage[utf8]{inputenc}
\usepackage[american]{babel}
\usepackage{graphicx,xcolor,bbold,titlesec}
\usepackage{braket}
\usepackage{MnSymbol}

\usepackage[breaklinks, pdftex, hyperfootnotes=true, pdfpagelabels, bookmarks, pageanchor]{hyperref}
\hypersetup{%
	colorlinks=true, linktocpage=true, pdfstartpage=1, pdfstartview=FitH, pdfborder={0 0 0},%
	breaklinks=true, pdfpagemode=UseNone, pageanchor=true, pdfpagemode=UseOutlines,%
	plainpages=false, bookmarksnumbered, bookmarksopen=true, bookmarksopenlevel=1,%
	hypertexnames=true, pdfhighlight=/O,
	urlcolor=red, linkcolor=red, citecolor=red,
	}
\usepackage{orcidlink}
\usepackage{tikz,ifthen}
\usepackage{bbold}
\usepackage{tikz-network}
\usetikzlibrary{patterns,decorations.pathreplacing,calligraphy}
\usetikzlibrary{shapes,arrows.meta,decorations.pathmorphing}

\newcommand{\Wg}{{\text{Wg}}} 

\newcommand{\Haar}{{\mathrm{Haar}}}
\newcommand{\murmps}{{\rm RMPS}} 
\newcommand{\murps}{{\rm RPS}} 

\newcommand{\Ex}{\mathbb{E}} 
\newcommand{\Var}{\text{Var}} 

\begin{document}

\newcommand{\titleinfo}{Anticoncentration and state design of random tensor networks}
\title{\titleinfo}

\author{Guglielmo Lami~\orcidlink{0000-0002-1778-7263}}
\affiliation{Laboratoire de Physique Th\'eorique et Mod\'elisation, CNRS UMR 8089,
CY Cergy Paris Universit\'e, 95302 Cergy-Pontoise Cedex, France}

\author{Jacopo De Nardis~\orcidlink{0000-0001-7877-0329}}
\affiliation{Laboratoire de Physique Th\'eorique et Mod\'elisation, CNRS UMR 8089,
CY Cergy Paris Universit\'e, 95302 Cergy-Pontoise Cedex, France}

\author{Xhek Turkeshi~\orcidlink{0000-0003-1093-3771}}
\affiliation{Institut für Theoretische Physik, Universität zu Köln, Zülpicher Strasse 77a, 50937 Köln, Germany}

\begin{abstract}
We investigate quantum random tensor network states in which the dimension of the bond scales polynomially with the size of the system $N$. Specifically, we examine the delocalization properties of random Matrix Product States (RMPS) in the computational basis by deriving an exact analytical expression for the Inverse Participation Ratio (IPR) of any degree, applicable to both open and closed boundary conditions. For bond dimensions $\chi \sim \gamma N$, we determine the leading order of the associated overlaps probability distribution and demonstrate its convergence to the Porter-Thomas distribution, characteristic of Haar-random states, as $\gamma$ increases. Additionally, we provide numerical evidence for the frame potential, measuring the $2$-distance from the Haar ensemble, which confirms the convergence of random MPS to Haar-like behavior for $\chi \gg  \sqrt{N}$. We extend this analysis to two-dimensional systems using random Projected Entangled Pair States (PEPS), where we similarly observe the convergence of IPRs to their Haar values for $\chi \gg \sqrt{N}$. These findings demonstrate that random tensor networks with bond dimensions scaling polynomially in the system size are fully Haar-anticoncentrated and approximate unitary designs, regardless of the spatial dimension.
\end{abstract}

\maketitle

\emph{Introduction. ---} 
Digital quantum many-body systems are a constant source of amazement, not only for their direct connection to quantum computing, but for their unique phenomenology, naturally captured by their quantum circuit description~\cite{Preskill_2018,Fisher2023}. 
Fundamental problems that can be effectively explored using quantum circuits are scrambling and ergodicity. 
When framed in terms of typical states generated by quantum circuits from a simple reference state, these questions reduce to the key concepts of anticoncentration and design.

Anticoncentration~\cite{dalzell2022random,eisert2023computational}, or Hilbert space delocalization~\cite{mace2019multifractal,sierant2022universal}, measures the extent of scrambling in a system by how widely the many-body wave function spreads across the computational basis. Highly anticoncentrated states are broadly distributed among basis elements, making them challenging to simulate or learn with classical computers.
Design is inherently tied to quantum randomness~\cite{choi2023preparing}. 
Typical quantum circuits make a many-body state look nearly indistinguishable from a uniformly random state in the Hilbert space. 
The ergodicity of a quantum system is then quantified by its ability to approximate the Haar distribution up to the $k$-th moment, a property known as $k$-designs~\cite{gross2007evenly,mele2024introduction,brandao2016local,ippoliti2022solvable,fava2024designsfreeprobability,cotler2023emergent,claeys2022emergentquantum}.

Despite their fundamental importance, including their role in benchmarking computational quantum advantage~\cite{boixo2018characterizing,arute2019quantum,2023Google} or their connection to thermalization~\cite{pappalardi2022eigenstate,fritzsch2024microcanonicalfreecumulantslattice,foini2024outofequilibriumeigenstatethermalizationhypothesis,pappalardi2024eigenstatethermalizationfreecumulants}, understanding anticoncentration and design in generic finite-depth circuits remains challenging. 
This difficulty arises from the complexity of calculating these quantities, which require resolving the intricate structure of the many-body state, often captured by non-linear indicators like participation entropy~\cite{luitz2014universal,luitz2014participation} or the frame potential~\cite{gross2021schur,mele2024introduction,Hunter2019,2206.14205,Tiutiakina2024}. 
Recent progress has been achieved in brickwork quantum circuits by mapping these indicators to statistical mechanical models, revealing a deep connection between anticoncentration and design~\cite{bertoni2024shallow,bertini2020scrambling,cioli2024approximateinversemeasurementchannel,turkeshi2024hilbert,mark2023benchmarking,mark2024federica,fefferman2024anticoncentration,christopoulos2024alexios,claeys2024fockspace}. 
However, the problem remains unresolved for more general systems, where intrinsic structural properties impose significant constraints on these behaviors. 

This letter resolves the anticoncentration and design properties of random tensor network states. These states, constrained by their bond dimension $\chi$, are not only pivotal in the study of holographic quantum gravity~\cite{hayden2016holographic,qi2017holographic,cheng2024random,piroli2020a}, but also play a crucial role in quantum many-body numerical techniques~\cite{schollwock2011the,biamonte2020lectures,silvi2019the,Orus_2014,Ranabhat_2022,Ranabhat_2024,Verstraete_2004,Jordan_2008,PhysRevLett.124.037201,Corboz_2011}. Tensor Network states are experimentally relevant as well, being readily prepared on current quantum platforms~\cite{piroli2021quantum,malz2024preparation,smith2024kevin,david2024preparing,zhang2024yifan}. 
Concretely, we focus on random matrix product states (RMPS) in one dimension~\cite{garnerone2010typicality,garnerone2010statistical,lancien2021correlation,haferkamp2021emergent,lami2024quantum,haag2023typical} and random projected entangled-pair states (PEPS) in two dimensions.
Our findings demonstrates that \emph{random tensor network with bond dimension $\chi$ polynomial in the system size $N$ form a compelling ensemble of random states}. In fact, despite exhibiting only sub-extensive entanglement, $S \sim \log N$, these states are fully anticoncentrated and can approximate Haar-random states with arbitrary precision. Given that random tensor networks can be prepared with finite depth $O(\log \chi)$ circuits \cite{smith2024kevin,david2024preparing,zhang2024yifan,Yuen2024}, they constitute a class of quantum states which well mimic Haar states and which can therefore be employed in different aspects of quantum simulations. 

The paper is organized as follows. After establishing general notations and outlining key methodological aspects, we explore the anticoncentration properties of RMPS by computing the exact Inverse Participation Ratio (IPR), which converges to the Haar value for $\chi\sim N$. 
This result allows us to determine the distribution of overlaps in the scaling limit, which approaches the Porter-Thomas distribution~\cite{porter1956,haake2001,Mullane2020,Hunter2019} for large value of the scaling parameter $\gamma \sim \chi/N$. 
We then compare these findings with the design properties of RMPS, numerically investigating the Frame Potential (FP) and observing convergence to the Haar value for $\chi \sim N^{1/2}$. Finally, we extend our analysis to two-dimensional PEPS, showing that for a system of $N = L^2$ qubits, convergence to Haar values occurs when $\chi \sim L$. 

\emph{Preliminaries and methods. ---} 
We consider a quantum system of $N$ qudits, with local and total Hilbert space dimension, denoted respectively as $d$ and $D=d^N$. 
We will employ inverse participation ratios to reveal the anticoncentration features of the states of interest tensor network, while the frame potential for their $k$-design properties. 
For a given state, $|\psi\rangle$ the IPR with respect to the computational basis $\mathcal{B} = \{ |\mathbf{x}\rangle\}_{x=0}^{D-1}$ is 
\begin{equation}
\mathcal{I}^{(k)}(\ket{\psi}) \equiv \sum_{\pmb{x}} |\langle \pmb{x}|\psi\rangle|^{2k} \, .
\end{equation}

\begin{figure}[t!]
\centering
\includegraphics[width=0.9\columnwidth]
{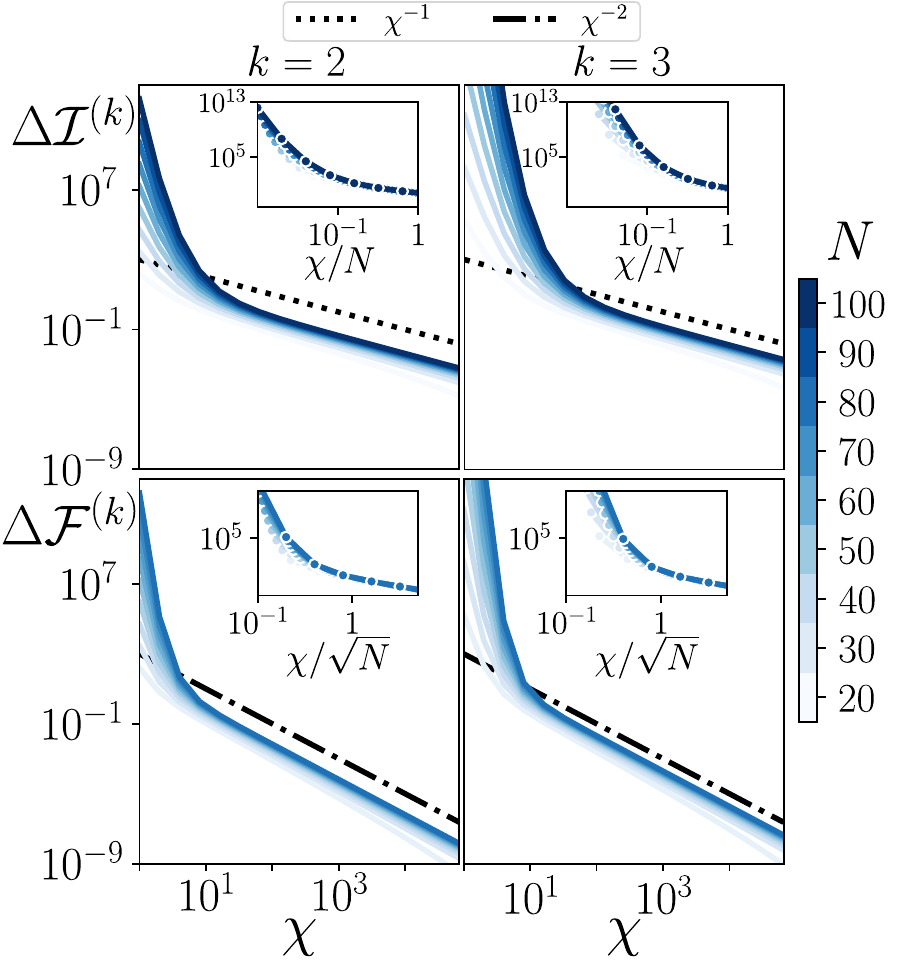}
\caption{\textit{Upper panels}: the IPR of RMPS $\Delta \mathcal{I}^{(k)} = \mathcal{I}_{\rm RMPS}^{(k)}/\mathcal{I}_{\rm Haar}^{(k)}-1$ obtained in Eq.~\eqref{eq:ipr_rnmps_obc}. \textit{Lower panels}: the frame potential of RMPS $\Delta \mathcal{F}^{(k)} = \mathcal{F}_{\rm RMPS}^{(k)}/\mathcal{F}_{\rm Haar}^{(k)}-1$ obtained by numerical contraction. We set $k=2,3$, $d=2$ and explore different $N$. Dotted and dashed lines indicate the decay with bond dimension $\chi$. Insets: same quantities are shown with $\chi$ rescaled by $N$ (IPR) and $\sqrt{N}$ (frame potential).}
\label{fig:ikfkrmps}
\end{figure}

Notice that $\mathcal{I}^{(1)}(\ket{\psi}) = 1$ is the normalization condition, while for $k=2$ one obtains the so-called collision probability~\cite{dalzell2022random}. We are interested in the average value of IPR over a given ensemble of states $\mathcal{E} = \{ \ket{\psi_i} \in \mathcal{H} \}_{i=1}^K$: 
\begin{equation}
\mathcal{I}^{(k)}_{\mathcal{E}} = \mathbb{E}_{\psi \sim \mathcal{E}}[\mathcal{I}^{(k)}(\ket{\psi})] = D \Ex_{\pmb{x} \sim \mathcal{B}, \psi \sim \mathcal{E}}[|\braket{ \pmb{x} | \psi}|^{2k}] \, ,
\end{equation}
where the ensemble average is defined as $\Ex_{\mathcal{E}} [\dots] = 1/K \sum_{i=1}^K (\dots)$. A more refined measure of anticoncentration involves the statistics of the overlap. Concretely, we define the random variable $w = D |\braket{\pmb{x} | \psi}|^{2}$, whose probability distribution is given by 
\begin{equation}\label{eq:p_iprd}
    \mathcal{P}_{\text{IPR}}(w) \equiv \Ex_{\pmb{x} \sim \mathcal{B}, \psi \sim \mathcal{E}} \big[ \delta \left(w - D |\braket{\pmb{x} | \psi}|^{2} \right) \big] \, .
\end{equation}
A simple computation reveals that the IPR corresponds to the $k$-th moment of $\mathcal{P}_{\text{IPR}}(w)$, up to multiplicative constant. As a result, the knowledge of $\mathcal{I}^{(k)}_{\mathcal{E}}$ for any $k$ allows to reconstruct the exact overlap distribution $\mathcal{P}_{\mathcal{E}}(w)$~\cite{turkeshi2023paulispectrummagictypical}. We will employ this remark in the following to study the deviation from the Porter-Thomas (PT) distribution 
\begin{equation}\label{eq:p_ipr}
    \mathcal{P}_{\text{IPR}}(w) = \frac{D - 1}{D} \left(1 - \frac{w}{D} \right)^{D-2} \simeq_{\lim D\gg 1} e^{-w} \, ,
\end{equation}
which describes $w$ for Haar random states.

The frame potential is defined by
\begin{equation}
    \mathcal{F}^{(k)}_{\mathcal{E}} \equiv \Ex_{\psi, \psi' \sim \mathcal{E}}[|\braket{\psi | \psi'}|^{2k}]\; ,
\end{equation}
and measures the 2-norm distance from the Haar distribution~\cite{gross2007evenly,mele2024introduction}. 
Using the unitary invariance of the Haar measure, it follows that $\mathcal{I}^{(k)}_{\Haar} = D \mathcal{F}^{(k)}_{\Haar}$, with $\mathcal{F}^{(k)}_{\Haar}=\binom{D+k-1}{k}^{-1} \simeq k! D^{-k}$, where the second equality is true for large $D$~\cite{turkeshi2023measuring}. Moreover, a distribution $\mathcal{P}_{\text{FP}}(w)$ for the overlaps $D |\braket{\psi | \psi'}|^{2}$ can be defined analogously to Eq.~\eqref{eq:p_iprd}, which generates the frame potentials.
In general, however, IPR and FP capture different aspects of random state ensembles, as we demonstrate below with concrete examples.

Methodologically, our calculations employ the Weingarten calculus, which we briefly revisit below in the vectorization formalism: $A\mapsto |A\rrangle$, with $\mathrm{tr}(A^\dagger B) = \llangle A|B\rrangle$ and $U^\dagger A U\mapsto (U^*\otimes U )|A\rrangle$~\cite{CHOI1975285,mele2024introduction}. 
We will frequently compute averages of $k$ copies of Haar-distributed matrices $U$ acting on a Hilbert space dimension of size $q$. 
Given the permutation operators on $k$ copies of the Hilbert space as $|T_\sigma\rrangle$ with $\sigma\in S_k$, which we graphically denote as a tensor with three indices $
\begin{tikzpicture}[baseline=(current  bounding  box.center), scale=0.5]
\tikzset{snake it/.style={decorate, decoration=snake}}
\draw[ultra thick, black] (0.5,0.6) -- (-0.5,0.6);
\draw[ultra thick, black] (0.5,0.4) -- (-0.5,0.4);
\draw [ultra thick, orange, snake it] (0.5,0.5) -- (2.0,0.5);
\draw[ultra thick, fill=white] (0,0) rectangle (1.,1.);
\node[scale=1.5] at (0.5, 0.5){\footnotesize $T$ \normalsize}; 
\end{tikzpicture} \;,$
we have 
\begin{equation}\label{eq:haark}
\begin{split}
    \Ex_{U \sim \Haar}&[(U^{*}  \otimes U)^{\otimes k}]
    =\sum_{\sigma, \pi \in S_k} \Wg_{\sigma,\pi}{(q)} |T_{\sigma}\rrangle  \llangle T_{\pi}| \, \\
    &=\raisebox{1.8ex}{\begin{tikzpicture}[baseline=(current  bounding  box.center), scale=0.55]
\tikzset{snake it/.style={decorate, decoration=snake}}
\pgfmathsetmacro{\ll}{3.5}
\draw[ultra thick, black] (0.5,0.6) -- (-0.5,0.6);
\draw[ultra thick, black] (0.5,0.4) -- (-0.5,0.4);
\draw[ultra thick, black] (\ll+0.5,0.6) -- (\ll+1.5,0.6);
\draw[ultra thick, black] (\ll+0.5,0.4) -- (\ll+1.5,0.4);
\draw [ultra thick, orange, snake it] (0.5,0.5) -- (4.0,0.5);
\draw[ultra thick, fill=white] (0,0) rectangle (1.,1.);
\draw[ultra thick, fill=white] (\ll,0) rectangle (\ll+1,1.);
\node[draw, diamond, fill=orange, ultra thick] (W) at (\ll/2+0.5, 0.5) {};
\node[scale=1.] at (\ll/2+0.5, 1.5) {$W(q)$};
\end{tikzpicture}}\;.
\end{split}
\end{equation}
where $W(q)$ is the Weingarten matrix $\Wg_{\sigma,\pi}{(q)}$~\cite{Kostenberger_2021}. 

\emph{Anticoncentration of Random Matrix Product States.} 
Matrix Product States (MPS) are defined by a list of tensors $A^{(i)}_{\alpha \beta}(x_i)$ ($i=1,2...,N$) where $\alpha, \beta \in \{1,2 ... ,\chi \}$ are indices that live in an auxiliary space of dimension $\chi$ and $x_i \in \{0,1, ...d-1\}$ is the local qudit variable~\cite{schollwock2011the}. 
The wave function $\psi_{\pmb{x}}=\braket{x_1, \dots,x_N | \psi}$ is obtained by contracting the $A^{(i)}$ along the auxiliary dimension. RMPS are defined by taking $A^{(i)}$ as isometries generated by applying a Haar unitary matrix $U^{(i)}$, of size $d \chi$, to the reference state $|0\rangle$ in the local computational basis~\cite{haferkamp2021emergent,haag2023typical}. 
Graphically, the wave function $\psi_{\pmb{x}}$ of a RMPS is given by
\begin{equation}\label{eq:rmps_bulk}
\begin{tikzpicture}[baseline=(current  bounding  box.center),scale=0.8]
\definecolor{mycolor}{rgb}{0.82,0.82,1.}
    \foreach \x in {1,...,5}{
        \draw[thick, black] (1.1*\x,0) -- (1.1*\x,1.6);
        \draw[thick, fill=white] (1.1*\x,0) circle (0.2);
        \node[scale=0.9] at (1.1*\x,-0.45) {$|0\rangle$};
    }
    \draw[line width=0.5mm, blue!60!black, dotted] (-0.3,0.8) -- (0.2,0.8);
    \draw[line width=0.5mm, blue!60!black] (0.2,0.8) -- (6.4,0.8);
    \draw[line width=0.5mm, blue!60!black, dotted] (6.4,0.8) -- (6.9,0.8);
    \foreach \x in {1,...,5}{
        \draw[thick, fill=mycolor] (1.1*\x-0.4,0.4) rectangle (1.1*\x+0.4,1.2);
        \pgfmathsetmacro{\y}{int(\x - 3)}
        \node[scale=0.6] at (1.1*\x,0.8) {$U^{(i \ifnum\y<0 
            \y 
        \else
            \ifnum\y=0
                \,
            \else
                - \y 
            \fi
        \fi)}$};
    }
\end{tikzpicture}\;.
\end{equation}

\begin{figure}[t!]
\centering
\includegraphics[width=\columnwidth]{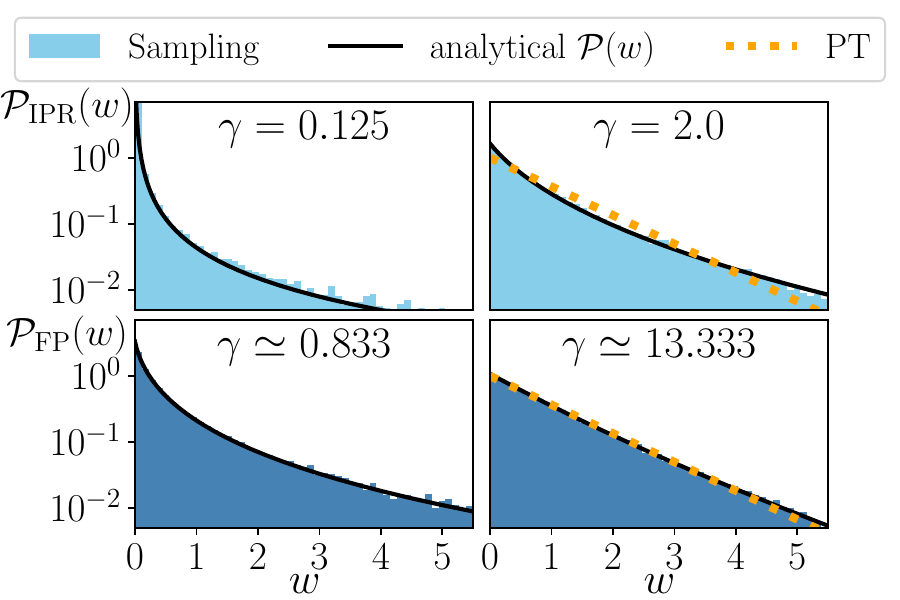}
\caption{Overlaps distributions of RMPS. \textit{Upper panels}: distribution over the computational basis $\mathcal{P}_{\text{IPR}}(w)$. \textit{Lower panels}: distribution of overlaps between RMPS $\mathcal{P}_{\text{FP}}(w)$. Black lines represent analytical expressions in the scaling limit (see Eqs.~\eqref{eq:scaling} and ~\eqref{eq:scaling_distribution}, and analogous relations for the FP).
The scaling parameter is defined as $\gamma = 2 \chi/N$ for the IPR, and 
$\gamma = a^{-1} \chi^2/N$ for the FP (with $a \simeq 0.6$). As $\gamma$ increases, the distributions converge towards the PT distribution (dotted orange line). We consider the case of qubits ($d=2$). Histograms are obtained by sampling random realizations of MPS ($N=32$, $\sim$ 50000 samples).
}
\label{fig:pwrmps}
\end{figure}

We are now in a position to outline the computation of the IPR $\mathcal{I}^{(k)}$ for RMPS. The strategy involves two steps. 
First, we notice that by unitary invariance of the Haar matrices $U^{(i)}$, we can always rotate the local computational basis states $\ket{x_i}$ to $\ket{0}$, yielding to $\mathcal{I}^{(k)}_{\murmps} = D \Ex_{\psi \sim \murmps}[|\braket{ \pmb{0} | \psi}|^{2k}]$. 
Contracting Eq.~\eqref{eq:rmps_bulk} with $|\pmb{0}\rangle$ and taking the average leads to $\mathcal{I}^{(k)}_{\murmps}$ fixed by only contractions of the transfer matrix 
\begin{equation}\label{eq:IPR_bulk}
\begin{tikzpicture}[baseline=(current bounding box.center), scale=1]
\tikzset{snake it/.style={decorate, decoration=snake}}
\node at (-0.9,0.6) {\Large $\mathcal{T}_\mathcal{I} \, \, = \, \, $};

\pgfmathsetmacro{\ll}{2.0}

\foreach \x in {1,...,1}{
    \node[draw, diamond, fill=blue, ultra thick] (G) at (2.5, 0.5) {};
    \node[scale=1.
] at (2.5, 1.0) {$G(\chi)$}  ;
    \node[scale=1.] at (1, 1.0) {$W(d\chi)$} ;
    \node[draw, diamond, fill=orange, ultra thick] (W) at (1, 0.5) {};
    \draw[ultra thick, orange, snake it] (W) -- (G);
}
\draw [ultra thick, orange, snake it] (-0.1,0.5) -- (W);
\draw [ultra thick, orange, snake it] (G) -- (3.7,0.5);

\end{tikzpicture} \;,
\end{equation}
where $G(\chi)$ is the Gram matrix with elements $G_{\sigma \pi}(\chi) = \llangle T_{\sigma} | T_{\pi} \rrangle = \chi^{\#(\sigma^{-1} \pi) }$, obtained by contracting permutation operators over $k$ replicas of a $\chi$-dimensional space, i.e.
\begin{equation}\label{eq:contraction0}
\raisebox{2.5ex}{
\begin{tikzpicture}[baseline=(current  bounding  box.center), scale=1.]
\tikzset{snake it/.style={decorate, decoration=snake}}
\pgfmathsetmacro{\ll}{1.5}
\draw [ultra thick, orange, snake it] (\ll/2-1,0.5) -- (\ll/2+1,0.5);
\node[draw, diamond, fill=blue, ultra thick] at (\ll/2, 0.5) {};
\node[scale=1.] at (\ll/2, 1.) {$G$};
\end{tikzpicture} }
{\text{\large =}}
\raisebox{0.6ex}{
\begin{tikzpicture}[baseline=(current  bounding  box.center), scale=0.7]
\pgfmathsetmacro{\ll}{2.5}
\tikzset{snake it/.style={decorate, decoration=snake}}
\draw [ultra thick, orange, snake it] (0.5,0.5) -- (-1.,0.5);
\draw [ultra thick, orange, snake it] (\ll+0.5,0.5) -- (\ll+2.,0.5);
\draw[ultra thick, black] (0.5,0.6) -- (\ll+0.5,0.6);
\draw[ultra thick, black] (0.5,0.4) -- (\ll+0.5,0.4);
\draw[ultra thick, fill=white] (0.1,0) rectangle (1.1,1.);
\draw[ultra thick, fill=white] (\ll-0.1,0) rectangle (\ll+1.-0.1,1.);
\node[scale=1.2] at (0.6, 0.5) {$T$};
\node[scale=1.2] at (\ll+0.4, 0.5) {$T$};
\end{tikzpicture} \;.
}
\end{equation}

\noindent The last ingredient is the contraction with the boundary condition. Throughout the Main Text we will consider open boundary conditions (OBC), but we detail the computation for periodic boundary conditions (PBC) in the End Matter. 
Similarly to Ref.~\cite{lami2024quantum}, the OBC computation involves contracting the transfer matrix with $|0\rrangle^{\otimes 2k}$, namely 
\begin{equation}
\begin{tikzpicture}[baseline=(current bounding box.center), scale=1]
\tikzset{snake it/.style={decorate, decoration=snake}}

\pgfmathsetmacro{\ll}{2.0}

\draw[line width=0.5mm, orange, dotted] (\ll/2*5,0.5) -- (\ll/2*5+0.3,0.5);

\foreach \x in {1,...,5}{
    \draw [ultra thick, orange, snake it] (\ll/2*\x-1, 0.5) -- (\ll/2*\x, 0.5);
}

\foreach \x in {1,...,2}{
    \node[draw, diamond, fill=blue, ultra thick] at (\ll*\x, 0.5) {};
    \node[scale=.9] at (\ll*\x, 1.0) {$G(\chi)$};
    \node[draw, diamond, fill=orange, ultra thick] at (\ll*\x-\ll/2, 0.5) {};
    \node[scale=.9] at (\ll*\x-\ll/2, 1.0) {$W(d\chi)$};
}

\draw[ultra thick, black] (\ll/2-1.,0.55) -- (\ll/2-1.9,0.55);
\draw[ultra thick, black] (\ll/2-1.,0.45) -- (\ll/2-1.9,0.45);
\draw[ultra thick, fill=white] (\ll/2-1.9, 0.5) circle (0.2);
\draw[ultra thick, fill=white] (\ll/2-1.2, 0.3) rectangle (\ll/2-0.8, 0.7);

\node[scale=1.] at (\ll/2-1., 1.0) {$T_\pi$};
\node[scale=1.] at (\ll/2-1.9, 1.0) {$\llangle 0|^{\otimes 2k}$};
\end{tikzpicture} \,.
\end{equation}
This expression can be contracted iteratively. We note that $(\llangle 0 |^{\otimes 2k})\cdot |T_\pi\rrangle = 1$ and recall the two identity concerning the Weingarten matrix 
\begin{equation}
    \sum_{\pi \in S_k} \Wg_{\pi,\tau}(q) = \frac{1}{q (q+1) ... (q+k-1)}\;,
\end{equation}
and the Gram matrix $\sum_{\pi \in S_k} G_{\pi,\tau}(q) = (q+1) ... (q+k-1)$.
Collecting all these remarks, we obtain the final expression
\begin{equation}\label{eq:ipr_rnmps_obc}
\begin{split}
   \mathcal{I}^{(k)}_{\murmps}  &= D \left(\frac{(\chi+1)\dots (\chi+k-1)}{d (d\chi+1)\dots (d\chi+k-1)}\right)^{N-r-1}\times \\ & \times \frac{k!}{d\chi (d\chi+1)\dots (d\chi+k-1)} \, ,
\end{split}
\end{equation}
with $r=\log_d \chi$. 
Notice that when $k=1$ one obtains the normalization condition $\mathcal{I}^{(k)}_{\murmps}=1$. Instead, when $\chi=d^{N-1}$ we obtain the exact Haar value $\mathcal{I}^{(k)}_{\Haar}  = D k! / \left( D(D+1)\dots (D+k-1) \right)$ as expected.

In Fig.~\ref{fig:ikfkrmps}, we plot Eq.~\eqref{eq:ipr_rnmps_obc} for $k=2,3$ and several values of $N$ as a function of $\chi$. In the scaling limit of large $\chi$, we expand our closed expression for $\mathcal{I}^{(k)}_{\murmps}$ to find the leading corrections to Haar
\begin{equation}\label{eq:IPRk}
\begin{split}  
\mathcal{I}^{(k)}_{\murmps} &\simeq \mathcal{I}^{(k)}_{\Haar} \left(1 + \frac{1}{\chi} \frac{k(k-1)}{2} \frac{d-1}{d}  \right)^N \, ,
\end{split}
\end{equation}
which is valid for any fixed $k$ and $d$ in the regime in which $N \ll \chi \ll D$. This formula is quite appealing. In a first place, it indicates that corrections to Haar scale down with $N/\chi$ (see dotted line in Figure~\ref{fig:ikfkrmps}). Secondly, it allows taking the thermodynamic limit $N \to \infty$, with $\chi =  N \gamma (d-1)/d$. In this limit, we find 
\begin{equation}\label{eq:scaling}
\lim_{\substack{N \to \infty \\ \chi =  N \gamma (d-1)/d}}  \mathcal{I}^{(k)}_{\murmps} = \mathcal{I}^{(k)}_{\Haar} e^{  k(k-1)/(2\gamma) } \, .
\end{equation}
The term $e^{k(k-1)/(2\gamma)}$ corresponds to the $k$th moment of the log-normal distribution. Therefore, in the scaling limit the overlaps $w$ are the product of two random variables $w= w_{\rm PT} w_{\rm LN}$, where the first is PT-distributed and the second is log-normally distributed. 
Hence, the distribution $\mathcal{P}_{\text{IPR}}$ of $w$ is expressed by the convolution
\begin{equation}\label{eq:scaling_distribution}
  \mathcal{P}(w;\gamma) =  \int \frac{du}{\sqrt{2\pi}}  \, e^{-u^2/2 + 1/\gamma}  \exp\left( -w e^{u/\sqrt{\gamma} + \frac{3}{2\gamma}} \right).
\end{equation}
Interestingly, a similar result was found recently in Ref.~\cite{christopoulos2024alexios} for the distribution of the overlaps in a random unitary circuit by keeping the ratio $x= N/N_{\rm th}(t)$, where $N_{\rm th}(t)$ is the Thouless length, to be constant and finite (see also Ref.~\cite{Chan2022}). In the present case of RMPS, the finite time and length of the circuit is replaced by the finite bond dimension $\chi$, which indeed can be considered as a cutoff in time and in the space correlations of the tensor network, given its finite correlation length scales polynomially in $\chi$ ~\cite{Loio2024Toappear}. 
In Figure~\ref{fig:pwrmps}, we plot the distribution obtained in Eq.~\eqref{eq:scaling_distribution} together with the histogram of overlaps obtained by a numerical sampling of random realizations of MPS~\cite{Stoudenmire_2010,Lami_2023_2,Lami_2024}. Remarkably, we find an excellent agreement, even for small values of $\gamma$, even if in the derivation of Eq.~\eqref{eq:scaling}, we assumed $N \ll \chi$, i.e.\ $d/(d-1) \ll \gamma$).

\emph{Frame potential of RMPS.---} The calculation of the FP for RMPS $\mathcal{F}^{(k)}_{\murmps} \equiv \Ex_{\psi, \psi' \sim \murmps}[|\braket{\psi | \psi'}|^{2k}]$ is more involved. In a standard brickwork quantum circuit, FP and IPR are equivalent up to a rescaling of time~\cite{christopoulos2024alexios}. Indeed, given two unitary brickwork circuits of depth $t$ $U_t$ and $U'_t$, one has $\langle 0 | (U'_t)^{\dag} U_t | 0 \rangle = \langle 0 | W_{2t} | 0 \rangle $, with $W_{2t}$ a new brickwork circuit of depth $2t$. However, this equivalence breaks down for RMPS due to their distinctive circuit structure. In fact, Eq.~\eqref{eq:rmps_bulk} can be viewed as a staircase quantum circuit consisting of sequential unitaries $U^{(i)}$ acting on $\log_d \chi + 1$ qudits, all initialized in $\ket{0}$~\cite{lami2024quantum,Lami_2023_1}. For illustrative purposes, when $r=2$, we have
\begin{equation}\label{eq:rmpsOBC}
   \begin{tikzpicture}[baseline=(current  bounding  box.center),scale=0.5]
   \definecolor{mycolor}{rgb}{0.82,0.82,1.}
    \foreach \x in {1,...,6}{
        \draw[thick, black] (\x,0) -- (\x,5);
        \draw[thick, fill=white] (\x,0) circle (0.2);
    }
    \foreach \x in {1,...,4}{
        \draw[thick, fill=mycolor] (\x-0.2,\x-0.4) rectangle (\x+2+0.2,\x+0.4);
        \node[scale=0.6] at (\x+1,\x) {$U^{( \ifnum\x<3 
            \x 
        \else
            \ifnum\x=3
                \, \dots
            \else
                N - r 
            \fi
        \fi)}$};
    }
\end{tikzpicture}\;.
\end{equation}
As a result, performing the averages and contracting the resulting network, we obtain the transfer matrix 
\begin{equation} \label{eq:frame_potential_bulk}
\begin{tikzpicture}[baseline=(current bounding box.center), scale=1]
\tikzset{snake it/.style={decorate, decoration=snake}}

\pgfmathsetmacro{\lv}{1.5}
\pgfmathsetmacro{\ll}{0.25}
\node at (-3.4,0.8) {\Large $\mathcal{T}_\mathcal{F} \, \, = \, \, $};

\draw[line width=0.5mm, orange, dotted] (-2.25,0) -- (-2,0);
\draw[line width=0.5mm, orange, dotted] (2.25,0) -- (2,0);
\draw[line width=0.5mm, orange, dotted] (-2.25,\lv) -- (-2,\lv);
\draw[line width=0.5mm, orange, dotted] (2.25,\lv) -- (2,\lv);

\draw [ultra thick, orange, snake it] (-2, 0.) -- (2, 0.);

\node[draw, diamond, fill=orange, ultra thick] at (-1, 0.) {};

\node[draw, diamond, fill=blue, ultra thick] at (1, 0.) {};

\draw [ultra thick, orange, snake it] (-2, \lv) -- (2, \lv);

\node[draw, diamond, fill=orange, ultra thick] at (-1, \lv) {};

\node[draw, diamond, fill=blue, ultra thick] at (1, \lv) {};

\draw[ultra thick, fill=orange, draw=white] (0,0) circle (0.13);

\draw[ultra thick, fill=orange, draw=white] (0,\lv) circle (0.13);

\draw [ultra thick, orange, snake it] (0, 0.) -- (0, \lv);

\draw[ultra thick, fill=white, draw=white] (-\ll/2,\lv/2-\lv/8-\ll/2) rectangle (\ll/2,\lv/2+\lv/8+\ll/2);

\draw[thick, black] (-\ll/4,\lv/2-\lv/8) -- (-\ll/4,\lv/2+\lv/8);
\draw[thick, black] (\ll/4,\lv/2-\lv/8) -- (\ll/4,\lv/2+\lv/8);

\draw[ultra thick, fill=white] (-\ll/2,\lv/2+\lv/8-\ll/2) rectangle (\ll/2,\lv/2+\lv/8+\ll/2);
\draw[ultra thick, fill=white] (-\ll/2,\lv/2-\lv/8-\ll/2) rectangle (\ll/2,\lv/2-\lv/8+\ll/2);
\end{tikzpicture}
\end{equation}
where the orange dot represents a copy tensor. Note that, compared to the simple form of the Eq.~\eqref{eq:IPR_bulk}, this does not lead to further analytical simplifications. Yet, these contractions are efficiently implementable numerically for small $k$. 
In Fig.~\ref{fig:ikfkrmps} (lower panels) we plot the deviation of the frame potential from Haar $\mathcal{F}^{(k)} / \mathcal{F}^{(k)}_{\Haar} - 1$ as a function of $\chi$, for $k=2,3$. These functions behave similarly to the IPR (upper panels), but with deviations decreasing as $N/\chi^{2}$ for large $\chi$, rather than $N/\chi$ as for the IPRs. In short, analogously to Eq.~\eqref{eq:IPRk}, we have 
\begin{equation}
   \mathcal{F}^{(k)}_{\murmps} = \mathcal{F}^{(k)}_{\rm Haar} \left( 1 + \frac{f(k,d)}{\chi^2} \right)^N \, ,
\end{equation}
where, by numerical inspection, we derive the dependence $f(k,d) = a k(k-1)/2$, mirroring the behavior found for IPRs. This suggests that the distribution of Eq.~\eqref{eq:scaling_distribution} also generates the FP, albeit in the different scaling regime $\chi = \sqrt{N \gamma a}  $. 
Intuitively, this different scaling is given by the fact  that, at least in the brickwork circuit, the frame potential at a time $t \sim \log \chi$ is equivalent to the IPR at the time $2t$,  corresponding to the transformation $\chi \to \chi^2$ in the RMPS circuit. Fig.~\ref{fig:pwrmps} presents a comparison between the analytical prediction~\eqref{eq:scaling_distribution} and sampling of RMPS. We use the value $a \simeq 0.6$ obtained by fitting the data of $\mathcal{F}^{(k)}_{\murmps}$ for $k=2,3,4$. The results show excellent agreement, even for small values of $\gamma$.

\begin{figure}[t!]
\centering
\includegraphics[width=0.9\columnwidth]{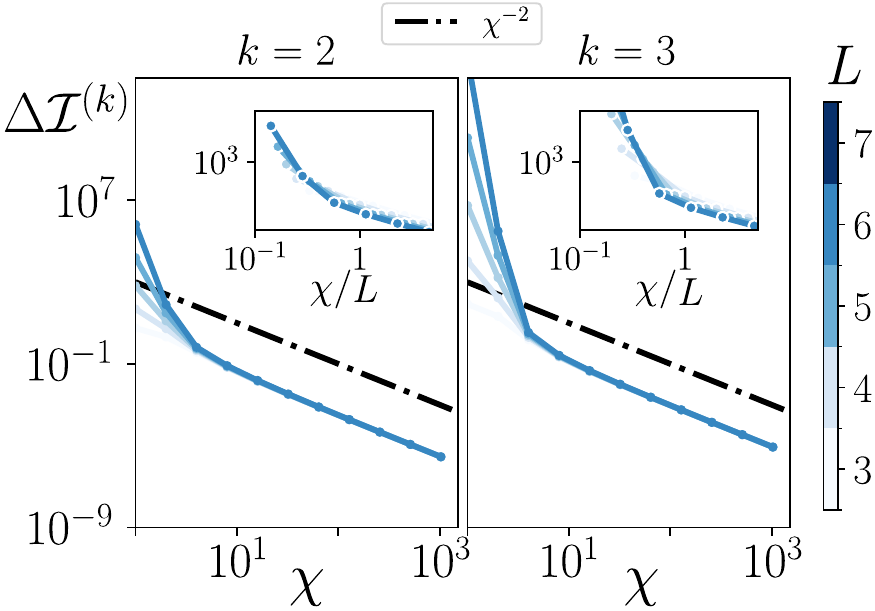}
\caption{The IPR of random PEPS $\Delta \mathcal{I}^{(k)} = \mathcal{I}_{\rm RMPS}^{(k)}/\mathcal{I}_{\rm Haar}^{(k)}-1$ obtained by numerical contraction. We set $k=2,3$, $d=2$ and explore different $L$. Insets: same quantities are shown with $\chi$ to by $L$ to show the crossover to the decay $\chi^{-2}$ for $\chi \ll L$.}
\label{fig:ik2Drmps}
\end{figure}

\emph{Anticoncentration of random PEPS. ---}
The extension of MPS to the 2D case is given by the Projected Entangled Pair States~\cite{Verstraete_2004,Jordan_2008,PhysRevLett.124.037201,Corboz_2011}. While in general PEPS cannot be constructed via a  sequential construction as the one used for MPS, here we focus on a subclass of PEPS in which all tensors are isometries~\cite{PhysRevLett.124.037201}.
For concreteness, we restrict our attention to a 2D lattice of qubits of size $N = L \times L$ and we consider the following random PEPS
\begin{equation}
\begin{tikzpicture}[baseline=(current  bounding  box.center),scale=0.8]
\pgfmathsetmacro{\lx}{0.25}
\pgfmathsetmacro{\ly}{0.15}
\definecolor{mycolor}{rgb}{0.82,0.82,1.}
    \foreach \x in {1,...,5}{
        \draw[line width=0.5mm, blue!60!black] (1.1*\x+5*\lx,+0.8+5*\ly) -- (1.1*\x+7*\lx,+0.8+7*\ly);
    }
    \foreach \y in {4,...,0}{
    \foreach \x in {1,...,5}{
        \draw[thick, black] (1.1*\x+\lx*\y, \ly*\y) -- (1.1*\x+\lx*\y, \ly*\y + 1.6);
        \draw[thick, fill=white] (1.1*\x+\lx*\y, \ly*\y) circle (0.2);
    }
    \draw[line width=0.5mm, blue!60!black, dotted] (\lx*\y-0.3,\ly*\y+0.8) -- (\lx*\y+0.2,\ly*\y+0.8);
    \draw[line width=0.5mm, blue!60!black] (0.2+\lx*\y,\ly*\y+0.8) -- (6.4+\lx*\y,\ly*\y+0.8);
    \draw[line width=0.5mm, blue!60!black, dotted] (6.4+\lx*\y,\ly*\y+0.8) -- (6.9+\lx*\y,\ly*\y+0.8);
    \foreach \x in {1,...,5}{
        \draw[thick, fill=mycolor] (1.1*\x-0.4+\lx*\y,\ly*\y+0.4) rectangle (1.1*\x+0.4+\lx*\y,\ly*\y+1.2);
    }
    }
    \foreach \x in {1,...,5}{
        \draw[line width=0.5mm, blue!60!black] (1.1*\x-2*\lx,+0.8-2*\ly) -- (1.1*\x,+0.8);
    }
\end{tikzpicture}\;,
\end{equation}
where each blue box is a unitary Haar matrix $U^{(i)}$ of size $d \chi^2$. To calculate the IPR, one can apply the same tricks we described in the 1D case. The result is a 2D tensor network in which the lattice sites carry the following fundamental block (transfer matrix):
\vspace{-2 mm}
\begin{equation}\label{eq:ipr_2d_bulk}
\begin{tikzpicture}[baseline=(current bounding box.center), scale=1]
\tikzset{snake it/.style={decorate, decoration=snake}}
\draw [ultra thick, orange, snake it] (-0.7, 0.) -- (0.7, 0.);
\draw [ultra thick, orange, snake it] (0., -0.7) -- (0, 0.7);
\draw[thick, fill=orange, draw=black] (0,0) circle (0.2);
\end{tikzpicture}
=
\raisebox{2.1ex}{
\begin{tikzpicture}[baseline=(current bounding box.center), scale=1]
\tikzset{snake it/.style={decorate, decoration=snake}}
\pgfmathsetmacro{\lv}{1.}
\pgfmathsetmacro{\ll}{0.25}

\draw [ultra thick, orange, snake it] (-3, 0.) -- (1.5, 0.);

\node[draw, diamond, fill=orange, ultra thick] at (-1, 0.) {};

\draw[ultra thick, fill=orange, draw=white] (0,0) circle (0.13);

\draw[ultra thick, fill=orange, draw=white] (-2,0) circle (0.13);

\draw [ultra thick, orange, snake it] (0, 0.) -- (0, 1.5);

\draw [ultra thick, orange, snake it] (-2, 0.) -- (-2, -1.);

\node[draw, diamond, fill=blue, ultra thick] at (0.75, 0.) {};

\node[draw, diamond, fill=blue, ultra thick] at (0., 0.75) {};
\end{tikzpicture}
}.
\end{equation}
The network can be efficiently contracted numerically, at least for small values of $L$, cf. Ref.~\cite{nahum2017quantum}. In Fig.~\ref{fig:ik2Drmps} we plot data obtained in this way for $L \in [3,7]$. 
We observe a crossover of IPRs for $\chi \sim L$ (see insets), after which the quantity $\mathcal{I}^{(k)} / \mathcal{I}^{(k)}_{\Haar} - 1$ decays as $\sim \chi^{-2}$, i.e. a faster decay in bond dimension compared to the 1D case.

\emph{Conclusion and outlooks.---} 
In this work we have studied the anticoncentration and state design properties of random tensor network states in one and two dimensions. 
Concretely, we have determined the exact inverse participation entropy of RMPS and compared these expressions with the frame potential obtained using robust numerical methods. Our findings highlight a substantial difference in how these properties approach the random (Haar) state predictions, with different scaling in the bond dimension. 
Remarkably, the IPR expressions allowed us to compute the exact distribution of overlaps between RMPS and computational basis states, providing a rare instance of fully analytical understanding of anticoncentration in finite-depth quantum circuits. 
Even rarer are the results we obtain for the two-dimensional geometry, where we successfully evaluated the inverse participation entropy using robust tensor network contraction.

Our findings directly connect with cross-entropy benchmarking (XEB), which has been used in the past years as a witness of quantum advantage.
For instance, Refs.~\cite{boixo2018characterizing,arute2019quantum,2023Google} investigated the statistical distribution of bit-string samples overlapped with the final state of a quantum circuit, and used the Porter-Thomas distribution as a benchmark. 
Our work offers key insights into the question of determining which classes of states exhibit full anticoncentration properties, as well as the cost of preparing them.
We have here shown that the \textit{Porter-Thomas distribution, and its finite-depth generalisation, can be achieved with tensor network states with only polynomial complexity}. We stress however that these states are not pseudoentangled in the sense defined in Refs.~\cite{AaronsonPseudo, IppolitiPseudo, GrewalPseudo}, since to make RMPS totally indistinguishable from Haar under a polynomial number of measurements would require to scale $\chi$ superpolynomially in $N$. The relationship between random tensor networks and pseudo-entanglement (and pseudo-magic~\cite{gu2024pseudomagic}) is a fertile future direction, see also~\cite{ippo}.

Our findings moreover align with recent research efforts aimed at identifying optimal tensor network states for quantum dynamics, particularly through the use of unitary Clifford gates to enhance their entanglement capacity~\cite{Paviglianiti_2024,lami2024quantum,MelloSantini2024,Qian2024,Qian20242,dowling2024magicheisenbergpicture}. Future research directions include the development of improved algorithms for approximating complex quantum states, generated by unitary evolution, with ensembles of MPS~\cite{PhysRevResearch.3.L022015,PhysRevLett.132.100402,PhysRevB.105.075131}, and further investigation into the anticoncentration of random tensor networks with $U(1)$~\cite{PhysRevX.8.031057,PhysRevX.8.031058} or non-abelian charges~\cite{PhysRevX.7.041046}. 

\begin{acknowledgments}
\paragraph{Acknowledgments. ---}  
We want to thank M. McGinley, W.W. Ho, S. Gopalakrishnan, M. Collura, L. Tagliacozzo, M. Ippoliti, L. Piroli, P. Sierant, and A. De Luca for inspiring discussions and for collaborations on topics connected with this work. J.D.N. and G.L. are founded by the ERC Starting Grant 101042293 (HEPIQ) and the ANR-22-CPJ1-0021-01. 
X.T. acknowledges support from DFG under Germany's Excellence Strategy – Cluster of Excellence Matter and Light for Quantum Computing (ML4Q) EXC 2004/1 – 390534769, and DFG Collaborative Research Center (CRC) 183 Project No. 277101999 - project B01. 
\end{acknowledgments}

\bibliographystyle{apsrev4-2}
\bibliography{bib}

\begin{thebibliography}{93}%
\makeatletter
\providecommand \@ifxundefined [1]{%
 \@ifx{#1\undefined}
}%
\providecommand \@ifnum [1]{%
 \ifnum #1\expandafter \@firstoftwo
 \else \expandafter \@secondoftwo
 \fi
}%
\providecommand \@ifx [1]{%
 \ifx #1\expandafter \@firstoftwo
 \else \expandafter \@secondoftwo
 \fi
}%
\providecommand \natexlab [1]{#1}%
\providecommand \enquote  [1]{``#1''}%
\providecommand \bibnamefont  [1]{#1}%
\providecommand \bibfnamefont [1]{#1}%
\providecommand \citenamefont [1]{#1}%
\providecommand \href@noop [0]{\@secondoftwo}%
\providecommand \href [0]{\begingroup \@sanitize@url \@href}%
\providecommand \@href[1]{\@@startlink{#1}\@@href}%
\providecommand \@@href[1]{\endgroup#1\@@endlink}%
\providecommand \@sanitize@url [0]{\catcode `\\12\catcode `\$12\catcode
  `\&12\catcode `\#12\catcode `\^12\catcode `\_12\catcode `\%12\relax}%
\providecommand \@@startlink[1]{}%
\providecommand \@@endlink[0]{}%
\providecommand \url  [0]{\begingroup\@sanitize@url \@url }%
\providecommand \@url [1]{\endgroup\@href {#1}{\urlprefix }}%
\providecommand \urlprefix  [0]{URL }%
\providecommand \Eprint [0]{\href }%
\providecommand \doibase [0]{https://doi.org/}%
\providecommand \selectlanguage [0]{\@gobble}%
\providecommand \bibinfo  [0]{\@secondoftwo}%
\providecommand \bibfield  [0]{\@secondoftwo}%
\providecommand \translation [1]{[#1]}%
\providecommand \BibitemOpen [0]{}%
\providecommand \bibitemStop [0]{}%
\providecommand \bibitemNoStop [0]{.\EOS\space}%
\providecommand \EOS [0]{\spacefactor3000\relax}%
\providecommand \BibitemShut  [1]{\csname bibitem#1\endcsname}%
\let\auto@bib@innerbib\@empty
\bibitem [{\citenamefont {Preskill}(2018)}]{Preskill_2018}%
  \BibitemOpen
  \bibfield  {author} {\bibinfo {author} {\bibfnamefont {J.}~\bibnamefont
  {Preskill}},\ }\href {https://doi.org/10.22331/q-2018-08-06-79} {\bibfield
  {journal} {\bibinfo  {journal} {{Quantum}}\ }\textbf {\bibinfo {volume}
  {2}},\ \bibinfo {pages} {79} (\bibinfo {year} {2018})}\BibitemShut {NoStop}%
\bibitem [{\citenamefont {Fisher}\ \emph {et~al.}(2023)\citenamefont {Fisher},
  \citenamefont {Khemani}, \citenamefont {Nahum},\ and\ \citenamefont
  {Vijay}}]{Fisher2023}%
  \BibitemOpen
  \bibfield  {author} {\bibinfo {author} {\bibfnamefont {M.~P.}\ \bibnamefont
  {Fisher}}, \bibinfo {author} {\bibfnamefont {V.}~\bibnamefont {Khemani}},
  \bibinfo {author} {\bibfnamefont {A.}~\bibnamefont {Nahum}},\ and\ \bibinfo
  {author} {\bibfnamefont {S.}~\bibnamefont {Vijay}},\ }\href
  {http://dx.doi.org/10.1146/annurev-conmatphys-031720-030658} {\bibfield
  {journal} {\bibinfo  {journal} {Ann. Rev. Cond. Mat. Phys.}\ }\textbf
  {\bibinfo {volume} {14}},\ \bibinfo {pages} {335} (\bibinfo {year}
  {2023})}\BibitemShut {NoStop}%
\bibitem [{\citenamefont {Dalzell}\ \emph {et~al.}(2022)\citenamefont
  {Dalzell}, \citenamefont {Hunter-Jones},\ and\ \citenamefont
  {Brand\~ao}}]{dalzell2022random}%
  \BibitemOpen
  \bibfield  {author} {\bibinfo {author} {\bibfnamefont {A.~M.}\ \bibnamefont
  {Dalzell}}, \bibinfo {author} {\bibfnamefont {N.}~\bibnamefont
  {Hunter-Jones}},\ and\ \bibinfo {author} {\bibfnamefont {F.~G. S.~L.}\
  \bibnamefont {Brand\~ao}},\ }\href
  {https://link.aps.org/doi/10.1103/PRXQuantum.3.010333} {\bibfield  {journal}
  {\bibinfo  {journal} {PRX Quantum}\ }\textbf {\bibinfo {volume} {3}},\
  \bibinfo {pages} {010333} (\bibinfo {year} {2022})}\BibitemShut {NoStop}%
\bibitem [{\citenamefont {Hangleiter}\ and\ \citenamefont
  {Eisert}(2023)}]{eisert2023computational}%
  \BibitemOpen
  \bibfield  {author} {\bibinfo {author} {\bibfnamefont {D.}~\bibnamefont
  {Hangleiter}}\ and\ \bibinfo {author} {\bibfnamefont {J.}~\bibnamefont
  {Eisert}},\ }\href {https://link.aps.org/doi/10.1103/RevModPhys.95.035001}
  {\bibfield  {journal} {\bibinfo  {journal} {Rev. Mod. Phys.}\ }\textbf
  {\bibinfo {volume} {95}},\ \bibinfo {pages} {035001} (\bibinfo {year}
  {2023})}\BibitemShut {NoStop}%
\bibitem [{\citenamefont {Mac\'e}\ \emph {et~al.}(2019)\citenamefont {Mac\'e},
  \citenamefont {Alet},\ and\ \citenamefont
  {Laflorencie}}]{mace2019multifractal}%
  \BibitemOpen
  \bibfield  {author} {\bibinfo {author} {\bibfnamefont {N.}~\bibnamefont
  {Mac\'e}}, \bibinfo {author} {\bibfnamefont {F.}~\bibnamefont {Alet}},\ and\
  \bibinfo {author} {\bibfnamefont {N.}~\bibnamefont {Laflorencie}},\ }\href
  {https://link.aps.org/doi/10.1103/PhysRevLett.123.180601} {\bibfield
  {journal} {\bibinfo  {journal} {Phys. Rev. Lett.}\ }\textbf {\bibinfo
  {volume} {123}},\ \bibinfo {pages} {180601} (\bibinfo {year}
  {2019})}\BibitemShut {NoStop}%
\bibitem [{\citenamefont {Sierant}\ and\ \citenamefont
  {Turkeshi}(2022)}]{sierant2022universal}%
  \BibitemOpen
  \bibfield  {author} {\bibinfo {author} {\bibfnamefont {P.}~\bibnamefont
  {Sierant}}\ and\ \bibinfo {author} {\bibfnamefont {X.}~\bibnamefont
  {Turkeshi}},\ }\href
  {https://link.aps.org/doi/10.1103/PhysRevLett.128.130605} {\bibfield
  {journal} {\bibinfo  {journal} {Phys. Rev. Lett.}\ }\textbf {\bibinfo
  {volume} {128}},\ \bibinfo {pages} {130605} (\bibinfo {year}
  {2022})}\BibitemShut {NoStop}%
\bibitem [{\citenamefont {Choi}\ \emph {et~al.}(2023)\citenamefont {Choi},
  \citenamefont {Shaw}, \citenamefont {Madjarov}, \citenamefont {Xie},
  \citenamefont {Finkelstein}, \citenamefont {Covey}, \citenamefont {Cotler},
  \citenamefont {Mark}, \citenamefont {Huang}, \citenamefont {Kale},
  \citenamefont {Pichler}, \citenamefont {Brandão}, \citenamefont {Choi},\
  and\ \citenamefont {Endres}}]{choi2023preparing}%
  \BibitemOpen
  \bibfield  {author} {\bibinfo {author} {\bibfnamefont {J.}~\bibnamefont
  {Choi}}, \bibinfo {author} {\bibfnamefont {A.~L.}\ \bibnamefont {Shaw}},
  \bibinfo {author} {\bibfnamefont {I.~S.}\ \bibnamefont {Madjarov}}, \bibinfo
  {author} {\bibfnamefont {X.}~\bibnamefont {Xie}}, \bibinfo {author}
  {\bibfnamefont {R.}~\bibnamefont {Finkelstein}}, \bibinfo {author}
  {\bibfnamefont {J.~P.}\ \bibnamefont {Covey}}, \bibinfo {author}
  {\bibfnamefont {J.~S.}\ \bibnamefont {Cotler}}, \bibinfo {author}
  {\bibfnamefont {D.~K.}\ \bibnamefont {Mark}}, \bibinfo {author}
  {\bibfnamefont {H.-Y.}\ \bibnamefont {Huang}}, \bibinfo {author}
  {\bibfnamefont {A.}~\bibnamefont {Kale}}, \bibinfo {author} {\bibfnamefont
  {H.}~\bibnamefont {Pichler}}, \bibinfo {author} {\bibfnamefont {F.~G. S.~L.}\
  \bibnamefont {Brandão}}, \bibinfo {author} {\bibfnamefont {S.}~\bibnamefont
  {Choi}},\ and\ \bibinfo {author} {\bibfnamefont {M.}~\bibnamefont {Endres}},\
  }\href {http://dx.doi.org/10.1038/s41586-022-05442-1} {\bibfield  {journal}
  {\bibinfo  {journal} {Nature}\ }\textbf {\bibinfo {volume} {613}},\ \bibinfo
  {pages} {468–473} (\bibinfo {year} {2023})}\BibitemShut {NoStop}%
\bibitem [{\citenamefont {Gross}\ \emph {et~al.}(2007)\citenamefont {Gross},
  \citenamefont {Audenaert},\ and\ \citenamefont {Eisert}}]{gross2007evenly}%
  \BibitemOpen
  \bibfield  {author} {\bibinfo {author} {\bibfnamefont {D.}~\bibnamefont
  {Gross}}, \bibinfo {author} {\bibfnamefont {K.}~\bibnamefont {Audenaert}},\
  and\ \bibinfo {author} {\bibfnamefont {J.}~\bibnamefont {Eisert}},\ }\href
  {http://dx.doi.org/10.1063/1.2716992} {\bibfield  {journal} {\bibinfo
  {journal} {J. Math. Phys.}\ }\textbf {\bibinfo {volume} {48}},\ \bibinfo
  {pages} {052104} (\bibinfo {year} {2007})}\BibitemShut {NoStop}%
\bibitem [{\citenamefont {Mele}(2024)}]{mele2024introduction}%
  \BibitemOpen
  \bibfield  {author} {\bibinfo {author} {\bibfnamefont {A.~A.}\ \bibnamefont
  {Mele}},\ }\href {https://doi.org/10.22331/q-2024-05-08-1340} {\bibfield
  {journal} {\bibinfo  {journal} {{Quantum}}\ }\textbf {\bibinfo {volume}
  {8}},\ \bibinfo {pages} {1340} (\bibinfo {year} {2024})}\BibitemShut
  {NoStop}%
\bibitem [{\citenamefont {Brandão}\ \emph {et~al.}(2016)\citenamefont
  {Brandão}, \citenamefont {Harrow},\ and\ \citenamefont
  {Horodecki}}]{brandao2016local}%
  \BibitemOpen
  \bibfield  {author} {\bibinfo {author} {\bibfnamefont {F.~G. S.~L.}\
  \bibnamefont {Brandão}}, \bibinfo {author} {\bibfnamefont {A.~W.}\
  \bibnamefont {Harrow}},\ and\ \bibinfo {author} {\bibfnamefont
  {M.}~\bibnamefont {Horodecki}},\ }\href
  {http://dx.doi.org/10.1007/s00220-016-2706-8} {\bibfield  {journal} {\bibinfo
   {journal} {Comm. Math. Phys.}\ }\textbf {\bibinfo {volume} {346}},\ \bibinfo
  {pages} {397–434} (\bibinfo {year} {2016})}\BibitemShut {NoStop}%
\bibitem [{\citenamefont {Ippoliti}\ and\ \citenamefont
  {Ho}(2022)}]{ippoliti2022solvable}%
  \BibitemOpen
  \bibfield  {author} {\bibinfo {author} {\bibfnamefont {M.}~\bibnamefont
  {Ippoliti}}\ and\ \bibinfo {author} {\bibfnamefont {W.~W.}\ \bibnamefont
  {Ho}},\ }\href {http://dx.doi.org/10.22331/q-2022-12-29-886} {\bibfield
  {journal} {\bibinfo  {journal} {Quantum}\ }\textbf {\bibinfo {volume} {6}},\
  \bibinfo {pages} {886} (\bibinfo {year} {2022})}\BibitemShut {NoStop}%
\bibitem [{\citenamefont {Fava}\ \emph {et~al.}()\citenamefont {Fava},
  \citenamefont {Kurchan},\ and\ \citenamefont
  {Pappalardi}}]{fava2024designsfreeprobability}%
  \BibitemOpen
  \bibfield  {author} {\bibinfo {author} {\bibfnamefont {M.}~\bibnamefont
  {Fava}}, \bibinfo {author} {\bibfnamefont {J.}~\bibnamefont {Kurchan}},\ and\
  \bibinfo {author} {\bibfnamefont {S.}~\bibnamefont {Pappalardi}},\
  }\href@noop {} {}\Eprint {https://arxiv.org/abs/2308.06200}
  {arXiv:2308.06200} \BibitemShut {NoStop}%
\bibitem [{\citenamefont {Cotler}\ \emph {et~al.}(2023)\citenamefont {Cotler},
  \citenamefont {Mark}, \citenamefont {Huang}, \citenamefont {Hern\'andez},
  \citenamefont {Choi}, \citenamefont {Shaw}, \citenamefont {Endres},\ and\
  \citenamefont {Choi}}]{cotler2023emergent}%
  \BibitemOpen
  \bibfield  {author} {\bibinfo {author} {\bibfnamefont {J.~S.}\ \bibnamefont
  {Cotler}}, \bibinfo {author} {\bibfnamefont {D.~K.}\ \bibnamefont {Mark}},
  \bibinfo {author} {\bibfnamefont {H.-Y.}\ \bibnamefont {Huang}}, \bibinfo
  {author} {\bibfnamefont {F.}~\bibnamefont {Hern\'andez}}, \bibinfo {author}
  {\bibfnamefont {J.}~\bibnamefont {Choi}}, \bibinfo {author} {\bibfnamefont
  {A.~L.}\ \bibnamefont {Shaw}}, \bibinfo {author} {\bibfnamefont
  {M.}~\bibnamefont {Endres}},\ and\ \bibinfo {author} {\bibfnamefont
  {S.}~\bibnamefont {Choi}},\ }\href
  {https://link.aps.org/doi/10.1103/PRXQuantum.4.010311} {\bibfield  {journal}
  {\bibinfo  {journal} {PRX Quantum}\ }\textbf {\bibinfo {volume} {4}},\
  \bibinfo {pages} {010311} (\bibinfo {year} {2023})}\BibitemShut {NoStop}%
\bibitem [{\citenamefont {Claeys}\ and\ \citenamefont
  {Lamacraft}(2022)}]{claeys2022emergentquantum}%
  \BibitemOpen
  \bibfield  {author} {\bibinfo {author} {\bibfnamefont {P.~W.}\ \bibnamefont
  {Claeys}}\ and\ \bibinfo {author} {\bibfnamefont {A.}~\bibnamefont
  {Lamacraft}},\ }\href {https://doi.org/10.22331/q-2022-06-15-738} {\bibfield
  {journal} {\bibinfo  {journal} {{Quantum}}\ }\textbf {\bibinfo {volume}
  {6}},\ \bibinfo {pages} {738} (\bibinfo {year} {2022})}\BibitemShut {NoStop}%
\bibitem [{\citenamefont {Boixo}\ \emph {et~al.}(2018)\citenamefont {Boixo},
  \citenamefont {Isakov}, \citenamefont {Smelyanskiy}, \citenamefont {Babbush},
  \citenamefont {Ding}, \citenamefont {Jiang}, \citenamefont {Bremner},
  \citenamefont {Martinis},\ and\ \citenamefont
  {Neven}}]{boixo2018characterizing}%
  \BibitemOpen
  \bibfield  {author} {\bibinfo {author} {\bibfnamefont {S.}~\bibnamefont
  {Boixo}}, \bibinfo {author} {\bibfnamefont {S.~V.}\ \bibnamefont {Isakov}},
  \bibinfo {author} {\bibfnamefont {V.~N.}\ \bibnamefont {Smelyanskiy}},
  \bibinfo {author} {\bibfnamefont {R.}~\bibnamefont {Babbush}}, \bibinfo
  {author} {\bibfnamefont {N.}~\bibnamefont {Ding}}, \bibinfo {author}
  {\bibfnamefont {Z.}~\bibnamefont {Jiang}}, \bibinfo {author} {\bibfnamefont
  {M.~J.}\ \bibnamefont {Bremner}}, \bibinfo {author} {\bibfnamefont {J.~M.}\
  \bibnamefont {Martinis}},\ and\ \bibinfo {author} {\bibfnamefont
  {H.}~\bibnamefont {Neven}},\ }\href
  {http://dx.doi.org/10.1038/s41567-018-0124-x} {\bibfield  {journal} {\bibinfo
   {journal} {Nature Phys.}\ }\textbf {\bibinfo {volume} {14}},\ \bibinfo
  {pages} {595–600} (\bibinfo {year} {2018})}\BibitemShut {NoStop}%
\bibitem [{\citenamefont {Arute}\ \emph {et~al.}(2019)\citenamefont {Arute},
  \citenamefont {Arya}, \citenamefont {Babbush}, \citenamefont {Bacon},
  \citenamefont {Bardin}, \citenamefont {Barends}, \citenamefont {Biswas},
  \citenamefont {Boixo}, \citenamefont {Brandao}, \citenamefont {Buell} \emph
  {et~al.}}]{arute2019quantum}%
  \BibitemOpen
  \bibfield  {author} {\bibinfo {author} {\bibfnamefont {F.}~\bibnamefont
  {Arute}}, \bibinfo {author} {\bibfnamefont {K.}~\bibnamefont {Arya}},
  \bibinfo {author} {\bibfnamefont {R.}~\bibnamefont {Babbush}}, \bibinfo
  {author} {\bibfnamefont {D.}~\bibnamefont {Bacon}}, \bibinfo {author}
  {\bibfnamefont {J.~C.}\ \bibnamefont {Bardin}}, \bibinfo {author}
  {\bibfnamefont {R.}~\bibnamefont {Barends}}, \bibinfo {author} {\bibfnamefont
  {R.}~\bibnamefont {Biswas}}, \bibinfo {author} {\bibfnamefont
  {S.}~\bibnamefont {Boixo}}, \bibinfo {author} {\bibfnamefont {F.~G.}\
  \bibnamefont {Brandao}}, \bibinfo {author} {\bibfnamefont {D.~A.}\
  \bibnamefont {Buell}}, \emph {et~al.},\ }\href
  {http://dx.doi.org/10.1038/s41586-019-1666-5} {\bibfield  {journal} {\bibinfo
   {journal} {Nature}\ }\textbf {\bibinfo {volume} {574}},\ \bibinfo {pages}
  {505–510} (\bibinfo {year} {2019})}\BibitemShut {NoStop}%
\bibitem [{\citenamefont {Acharya}\ and\ \citenamefont {{et
  al.}}(2023)}]{2023Google}%
  \BibitemOpen
  \bibfield  {author} {\bibinfo {author} {\bibfnamefont {R.}~\bibnamefont
  {Acharya}}\ and\ \bibinfo {author} {\bibnamefont {{et al.}}},\ }\href
  {http://dx.doi.org/10.1038/s41586-022-05434-1} {\bibfield  {journal}
  {\bibinfo  {journal} {Nature}\ }\textbf {\bibinfo {volume} {614}},\ \bibinfo
  {pages} {676–681} (\bibinfo {year} {2023})}\BibitemShut {NoStop}%
\bibitem [{\citenamefont {Pappalardi}\ \emph {et~al.}(2022)\citenamefont
  {Pappalardi}, \citenamefont {Foini},\ and\ \citenamefont
  {Kurchan}}]{pappalardi2022eigenstate}%
  \BibitemOpen
  \bibfield  {author} {\bibinfo {author} {\bibfnamefont {S.}~\bibnamefont
  {Pappalardi}}, \bibinfo {author} {\bibfnamefont {L.}~\bibnamefont {Foini}},\
  and\ \bibinfo {author} {\bibfnamefont {J.}~\bibnamefont {Kurchan}},\ }\href
  {https://link.aps.org/doi/10.1103/PhysRevLett.129.170603} {\bibfield
  {journal} {\bibinfo  {journal} {Phys. Rev. Lett.}\ }\textbf {\bibinfo
  {volume} {129}},\ \bibinfo {pages} {170603} (\bibinfo {year}
  {2022})}\BibitemShut {NoStop}%
\bibitem [{\citenamefont {Fritzsch}\ \emph {et~al.}()\citenamefont {Fritzsch},
  \citenamefont {Prosen},\ and\ \citenamefont
  {Pappalardi}}]{fritzsch2024microcanonicalfreecumulantslattice}%
  \BibitemOpen
  \bibfield  {author} {\bibinfo {author} {\bibfnamefont {F.}~\bibnamefont
  {Fritzsch}}, \bibinfo {author} {\bibfnamefont {T.}~\bibnamefont {Prosen}},\
  and\ \bibinfo {author} {\bibfnamefont {S.}~\bibnamefont {Pappalardi}},\
  }\href {https://arxiv.org/abs/2409.01404} {}\Eprint
  {https://arxiv.org/abs/2409.01404} {arXiv:2409.01404} \BibitemShut {NoStop}%
\bibitem [{\citenamefont {Foini}\ \emph {et~al.}()\citenamefont {Foini},
  \citenamefont {Dymarski},\ and\ \citenamefont
  {Pappalardi}}]{foini2024outofequilibriumeigenstatethermalizationhypothesis}%
  \BibitemOpen
  \bibfield  {author} {\bibinfo {author} {\bibfnamefont {L.}~\bibnamefont
  {Foini}}, \bibinfo {author} {\bibfnamefont {A.}~\bibnamefont {Dymarski}},\
  and\ \bibinfo {author} {\bibfnamefont {S.}~\bibnamefont {Pappalardi}},\
  }\href@noop {} {}\Eprint {https://arxiv.org/abs/2406.04684}
  {arXiv:2406.04684} \BibitemShut {NoStop}%
\bibitem [{\citenamefont {Pappalardi}\ \emph {et~al.}()\citenamefont
  {Pappalardi}, \citenamefont {Fritzsch},\ and\ \citenamefont
  {Prosen}}]{pappalardi2024eigenstatethermalizationfreecumulants}%
  \BibitemOpen
  \bibfield  {author} {\bibinfo {author} {\bibfnamefont {S.}~\bibnamefont
  {Pappalardi}}, \bibinfo {author} {\bibfnamefont {F.}~\bibnamefont
  {Fritzsch}},\ and\ \bibinfo {author} {\bibfnamefont {T.}~\bibnamefont
  {Prosen}},\ }\href@noop {} {}\Eprint {https://arxiv.org/abs/2303.00713}
  {arXiv:2303.00713} \BibitemShut {NoStop}%
\bibitem [{\citenamefont {Luitz}\ \emph
  {et~al.}(2014{\natexlab{a}})\citenamefont {Luitz}, \citenamefont {Alet},\
  and\ \citenamefont {Laflorencie}}]{luitz2014universal}%
  \BibitemOpen
  \bibfield  {author} {\bibinfo {author} {\bibfnamefont {D.~J.}\ \bibnamefont
  {Luitz}}, \bibinfo {author} {\bibfnamefont {F.}~\bibnamefont {Alet}},\ and\
  \bibinfo {author} {\bibfnamefont {N.}~\bibnamefont {Laflorencie}},\ }\href
  {https://link.aps.org/doi/10.1103/PhysRevLett.112.057203} {\bibfield
  {journal} {\bibinfo  {journal} {Phys. Rev. Lett.}\ }\textbf {\bibinfo
  {volume} {112}},\ \bibinfo {pages} {057203} (\bibinfo {year}
  {2014}{\natexlab{a}})}\BibitemShut {NoStop}%
\bibitem [{\citenamefont {Luitz}\ \emph
  {et~al.}(2014{\natexlab{b}})\citenamefont {Luitz}, \citenamefont
  {Laflorencie},\ and\ \citenamefont {Alet}}]{luitz2014participation}%
  \BibitemOpen
  \bibfield  {author} {\bibinfo {author} {\bibfnamefont {D.~J.}\ \bibnamefont
  {Luitz}}, \bibinfo {author} {\bibfnamefont {N.}~\bibnamefont {Laflorencie}},\
  and\ \bibinfo {author} {\bibfnamefont {F.}~\bibnamefont {Alet}},\ }\href
  {http://dx.doi.org/10.1088/1742-5468/2014/08/P08007} {\bibfield  {journal}
  {\bibinfo  {journal} {J. Stat. Mech.: Theor. Exp.}\ }\textbf {\bibinfo
  {volume} {2014}},\ \bibinfo {pages} {P08007} (\bibinfo {year}
  {2014}{\natexlab{b}})}\BibitemShut {NoStop}%
\bibitem [{\citenamefont {Gross}\ \emph {et~al.}(2021)\citenamefont {Gross},
  \citenamefont {Nezami},\ and\ \citenamefont {Walter}}]{gross2021schur}%
  \BibitemOpen
  \bibfield  {author} {\bibinfo {author} {\bibfnamefont {D.}~\bibnamefont
  {Gross}}, \bibinfo {author} {\bibfnamefont {S.}~\bibnamefont {Nezami}},\ and\
  \bibinfo {author} {\bibfnamefont {M.}~\bibnamefont {Walter}},\ }\href
  {https://doi.org/10.1007/s00220-021-04118-7} {\bibfield  {journal} {\bibinfo
  {journal} {Comm. Math. Phys.}\ }\textbf {\bibinfo {volume} {385}},\ \bibinfo
  {pages} {1325} (\bibinfo {year} {2021})}\BibitemShut {NoStop}%
\bibitem [{\citenamefont {Hunter-Jones}()}]{Hunter2019}%
  \BibitemOpen
  \bibfield  {author} {\bibinfo {author} {\bibfnamefont {N.}~\bibnamefont
  {Hunter-Jones}},\ }\href@noop {} {}\Eprint {https://arxiv.org/abs/1905.12053}
  {arXiv:1905.12053} \BibitemShut {NoStop}%
\bibitem [{\citenamefont {Jian}\ \emph {et~al.}()\citenamefont {Jian},
  \citenamefont {Bentsen},\ and\ \citenamefont {Swingle}}]{2206.14205}%
  \BibitemOpen
  \bibfield  {author} {\bibinfo {author} {\bibfnamefont {S.-K.}\ \bibnamefont
  {Jian}}, \bibinfo {author} {\bibfnamefont {G.}~\bibnamefont {Bentsen}},\ and\
  \bibinfo {author} {\bibfnamefont {B.}~\bibnamefont {Swingle}},\ }\href@noop
  {} {}\Eprint {https://arxiv.org/abs/2206.14205} {arXiv:2206.14205}
  \BibitemShut {NoStop}%
\bibitem [{\citenamefont {Tiutiakina}\ \emph {et~al.}(2024)\citenamefont
  {Tiutiakina}, \citenamefont {De~Luca},\ and\ \citenamefont
  {De~Nardis}}]{Tiutiakina2024}%
  \BibitemOpen
  \bibfield  {author} {\bibinfo {author} {\bibfnamefont {A.}~\bibnamefont
  {Tiutiakina}}, \bibinfo {author} {\bibfnamefont {A.}~\bibnamefont
  {De~Luca}},\ and\ \bibinfo {author} {\bibfnamefont {J.}~\bibnamefont
  {De~Nardis}},\ }\href {http://dx.doi.org/10.1007/JHEP01(2024)115} {\bibfield
  {journal} {\bibinfo  {journal} {J. High Energy Phys.}\ }\textbf {\bibinfo
  {volume} {2024}},\ \bibinfo {pages} {115}}\BibitemShut {NoStop}%
\bibitem [{\citenamefont {Bertoni}\ \emph {et~al.}(2024)\citenamefont
  {Bertoni}, \citenamefont {Haferkamp}, \citenamefont {Hinsche}, \citenamefont
  {Ioannou}, \citenamefont {Eisert},\ and\ \citenamefont
  {Pashayan}}]{bertoni2024shallow}%
  \BibitemOpen
  \bibfield  {author} {\bibinfo {author} {\bibfnamefont {C.}~\bibnamefont
  {Bertoni}}, \bibinfo {author} {\bibfnamefont {J.}~\bibnamefont {Haferkamp}},
  \bibinfo {author} {\bibfnamefont {M.}~\bibnamefont {Hinsche}}, \bibinfo
  {author} {\bibfnamefont {M.}~\bibnamefont {Ioannou}}, \bibinfo {author}
  {\bibfnamefont {J.}~\bibnamefont {Eisert}},\ and\ \bibinfo {author}
  {\bibfnamefont {H.}~\bibnamefont {Pashayan}},\ }\href
  {https://link.aps.org/doi/10.1103/PhysRevLett.133.020602} {\bibfield
  {journal} {\bibinfo  {journal} {Phys. Rev. Lett.}\ }\textbf {\bibinfo
  {volume} {133}},\ \bibinfo {pages} {020602} (\bibinfo {year}
  {2024})}\BibitemShut {NoStop}%
\bibitem [{\citenamefont {Bertini}\ and\ \citenamefont
  {Piroli}(2020)}]{bertini2020scrambling}%
  \BibitemOpen
  \bibfield  {author} {\bibinfo {author} {\bibfnamefont {B.}~\bibnamefont
  {Bertini}}\ and\ \bibinfo {author} {\bibfnamefont {L.}~\bibnamefont
  {Piroli}},\ }\href {https://link.aps.org/doi/10.1103/PhysRevB.102.064305}
  {\bibfield  {journal} {\bibinfo  {journal} {Phys. Rev. B}\ }\textbf {\bibinfo
  {volume} {102}},\ \bibinfo {pages} {064305} (\bibinfo {year}
  {2020})}\BibitemShut {NoStop}%
\bibitem [{\citenamefont {Cioli}\ \emph {et~al.}()\citenamefont {Cioli},
  \citenamefont {Ercolessi}, \citenamefont {Ippoliti}, \citenamefont
  {Turkeshi},\ and\ \citenamefont
  {Piroli}}]{cioli2024approximateinversemeasurementchannel}%
  \BibitemOpen
  \bibfield  {author} {\bibinfo {author} {\bibfnamefont {R.}~\bibnamefont
  {Cioli}}, \bibinfo {author} {\bibfnamefont {E.}~\bibnamefont {Ercolessi}},
  \bibinfo {author} {\bibfnamefont {M.}~\bibnamefont {Ippoliti}}, \bibinfo
  {author} {\bibfnamefont {X.}~\bibnamefont {Turkeshi}},\ and\ \bibinfo
  {author} {\bibfnamefont {L.}~\bibnamefont {Piroli}},\ }\href@noop {}
  {}\Eprint {https://arxiv.org/abs/2407.11813} {arXiv:2407.11813} \BibitemShut
  {NoStop}%
\bibitem [{\citenamefont {Turkeshi}\ and\ \citenamefont
  {Sierant}(2024)}]{turkeshi2024hilbert}%
  \BibitemOpen
  \bibfield  {author} {\bibinfo {author} {\bibfnamefont {X.}~\bibnamefont
  {Turkeshi}}\ and\ \bibinfo {author} {\bibfnamefont {P.}~\bibnamefont
  {Sierant}},\ }\href {https://doi.org/10.3390/e26060471} {\bibfield  {journal}
  {\bibinfo  {journal} {Entropy}\ }\textbf {\bibinfo {volume} {26}},\ \bibinfo
  {pages} {471} (\bibinfo {year} {2024})}\BibitemShut {NoStop}%
\bibitem [{\citenamefont {Mark}\ \emph {et~al.}(2023)\citenamefont {Mark},
  \citenamefont {Choi}, \citenamefont {Shaw}, \citenamefont {Endres},\ and\
  \citenamefont {Choi}}]{mark2023benchmarking}%
  \BibitemOpen
  \bibfield  {author} {\bibinfo {author} {\bibfnamefont {D.~K.}\ \bibnamefont
  {Mark}}, \bibinfo {author} {\bibfnamefont {J.}~\bibnamefont {Choi}}, \bibinfo
  {author} {\bibfnamefont {A.~L.}\ \bibnamefont {Shaw}}, \bibinfo {author}
  {\bibfnamefont {M.}~\bibnamefont {Endres}},\ and\ \bibinfo {author}
  {\bibfnamefont {S.}~\bibnamefont {Choi}},\ }\href
  {https://link.aps.org/doi/10.1103/PhysRevLett.131.110601} {\bibfield
  {journal} {\bibinfo  {journal} {Phys. Rev. Lett.}\ }\textbf {\bibinfo
  {volume} {131}},\ \bibinfo {pages} {110601} (\bibinfo {year}
  {2023})}\BibitemShut {NoStop}%
\bibitem [{\citenamefont {Mark}\ \emph {et~al.}()\citenamefont {Mark},
  \citenamefont {Surace}, \citenamefont {Elben}, \citenamefont {Shaw},
  \citenamefont {Choi}, \citenamefont {Refael}, \citenamefont {Endres},\ and\
  \citenamefont {Choi}}]{mark2024federica}%
  \BibitemOpen
  \bibfield  {author} {\bibinfo {author} {\bibfnamefont {D.~K.}\ \bibnamefont
  {Mark}}, \bibinfo {author} {\bibfnamefont {F.}~\bibnamefont {Surace}},
  \bibinfo {author} {\bibfnamefont {A.}~\bibnamefont {Elben}}, \bibinfo
  {author} {\bibfnamefont {A.~L.}\ \bibnamefont {Shaw}}, \bibinfo {author}
  {\bibfnamefont {J.}~\bibnamefont {Choi}}, \bibinfo {author} {\bibfnamefont
  {G.}~\bibnamefont {Refael}}, \bibinfo {author} {\bibfnamefont
  {M.}~\bibnamefont {Endres}},\ and\ \bibinfo {author} {\bibfnamefont
  {S.}~\bibnamefont {Choi}},\ }\href@noop {} {}\Eprint
  {https://arxiv.org/abs/2403.11970} {arXiv:2403.11970} \BibitemShut {NoStop}%
\bibitem [{\citenamefont {Fefferman}\ \emph {et~al.}()\citenamefont
  {Fefferman}, \citenamefont {Ghosh},\ and\ \citenamefont
  {Zhan}}]{fefferman2024anticoncentration}%
  \BibitemOpen
  \bibfield  {author} {\bibinfo {author} {\bibfnamefont {B.}~\bibnamefont
  {Fefferman}}, \bibinfo {author} {\bibfnamefont {S.}~\bibnamefont {Ghosh}},\
  and\ \bibinfo {author} {\bibfnamefont {W.}~\bibnamefont {Zhan}},\ }\href@noop
  {} {}\Eprint {https://arxiv.org/abs/2407.19561} {arXiv:2407.19561}
  \BibitemShut {NoStop}%
\bibitem [{\citenamefont {Christopoulos}\ \emph {et~al.}()\citenamefont
  {Christopoulos}, \citenamefont {Chan},\ and\ \citenamefont
  {Luca}}]{christopoulos2024alexios}%
  \BibitemOpen
  \bibfield  {author} {\bibinfo {author} {\bibfnamefont {A.}~\bibnamefont
  {Christopoulos}}, \bibinfo {author} {\bibfnamefont {A.}~\bibnamefont
  {Chan}},\ and\ \bibinfo {author} {\bibfnamefont {A.~D.}\ \bibnamefont
  {Luca}},\ }\href@noop {} {}\Eprint {https://arxiv.org/abs/2404.10057}
  {arXiv:2404.10057} \BibitemShut {NoStop}%
\bibitem [{\citenamefont {Claeys}\ and\ \citenamefont
  {Tomasi}()}]{claeys2024fockspace}%
  \BibitemOpen
  \bibfield  {author} {\bibinfo {author} {\bibfnamefont {P.~W.}\ \bibnamefont
  {Claeys}}\ and\ \bibinfo {author} {\bibfnamefont {G.~D.}\ \bibnamefont
  {Tomasi}},\ }\href@noop {} {}\Eprint {https://arxiv.org/abs/2408.02732}
  {arXiv:2408.02732} \BibitemShut {NoStop}%
\bibitem [{\citenamefont {Hayden}\ \emph {et~al.}(2016)\citenamefont {Hayden},
  \citenamefont {Nezami}, \citenamefont {Qi}, \citenamefont {Thomas},
  \citenamefont {Walter},\ and\ \citenamefont {Yang}}]{hayden2016holographic}%
  \BibitemOpen
  \bibfield  {author} {\bibinfo {author} {\bibfnamefont {P.}~\bibnamefont
  {Hayden}}, \bibinfo {author} {\bibfnamefont {S.}~\bibnamefont {Nezami}},
  \bibinfo {author} {\bibfnamefont {X.-L.}\ \bibnamefont {Qi}}, \bibinfo
  {author} {\bibfnamefont {N.}~\bibnamefont {Thomas}}, \bibinfo {author}
  {\bibfnamefont {M.}~\bibnamefont {Walter}},\ and\ \bibinfo {author}
  {\bibfnamefont {Z.}~\bibnamefont {Yang}},\ }\href
  {http://dx.doi.org/10.1007/JHEP11(2016)009} {\bibfield  {journal} {\bibinfo
  {journal} {J. High Energy Phys.}\ }\textbf {\bibinfo {volume} {2016}},\
  \bibinfo {pages} {9}}\BibitemShut {NoStop}%
\bibitem [{\citenamefont {Qi}\ \emph {et~al.}(2017)\citenamefont {Qi},
  \citenamefont {Yang},\ and\ \citenamefont {You}}]{qi2017holographic}%
  \BibitemOpen
  \bibfield  {author} {\bibinfo {author} {\bibfnamefont {X.-L.}\ \bibnamefont
  {Qi}}, \bibinfo {author} {\bibfnamefont {Z.}~\bibnamefont {Yang}},\ and\
  \bibinfo {author} {\bibfnamefont {Y.-Z.}\ \bibnamefont {You}},\ }\href
  {http://dx.doi.org/10.1007/JHEP08(2017)060} {\bibfield  {journal} {\bibinfo
  {journal} {J. High Energy Phys.}\ }\textbf {\bibinfo {volume} {2017}}\bibinfo
   {number} { (8)}}\BibitemShut {NoStop}%
\bibitem [{\citenamefont {Cheng}\ \emph {et~al.}({\natexlab{a}})\citenamefont
  {Cheng}, \citenamefont {Lancien}, \citenamefont {Penington}, \citenamefont
  {Walter},\ and\ \citenamefont {Witteveen}}]{cheng2024random}%
  \BibitemOpen
\bibfield  {number} {  }\bibfield  {author} {\bibinfo {author} {\bibfnamefont
  {N.}~\bibnamefont {Cheng}}, \bibinfo {author} {\bibfnamefont
  {C.}~\bibnamefont {Lancien}}, \bibinfo {author} {\bibfnamefont
  {G.}~\bibnamefont {Penington}}, \bibinfo {author} {\bibfnamefont
  {M.}~\bibnamefont {Walter}},\ and\ \bibinfo {author} {\bibfnamefont
  {F.}~\bibnamefont {Witteveen}},\ }\href@noop {} {} ({\natexlab{a}}),\ \Eprint
  {https://arxiv.org/abs/2206.10482} {arXiv:2206.10482} \BibitemShut {NoStop}%
\bibitem [{\citenamefont {Piroli}\ \emph {et~al.}(2020)\citenamefont {Piroli},
  \citenamefont {Sünderhauf},\ and\ \citenamefont {Qi}}]{piroli2020a}%
  \BibitemOpen
  \bibfield  {author} {\bibinfo {author} {\bibfnamefont {L.}~\bibnamefont
  {Piroli}}, \bibinfo {author} {\bibfnamefont {C.}~\bibnamefont
  {Sünderhauf}},\ and\ \bibinfo {author} {\bibfnamefont {X.-L.}\ \bibnamefont
  {Qi}},\ }\href {http://dx.doi.org/10.1007/JHEP04(2020)063} {\bibfield
  {journal} {\bibinfo  {journal} {J. High Energy Phys.}\ }\textbf {\bibinfo
  {volume} {2020}}\bibinfo  {number} { (4)},\ \bibinfo {pages}
  {63}}\BibitemShut {NoStop}%
\bibitem [{\citenamefont {Schollw\"ock}(2011)}]{schollwock2011the}%
  \BibitemOpen
\bibfield  {number} {  }\bibfield  {author} {\bibinfo {author} {\bibfnamefont
  {U.}~\bibnamefont {Schollw\"ock}},\ }\href
  {https://www.sciencedirect.com/science/article/pii/S0003491610001752}
  {\bibfield  {journal} {\bibinfo  {journal} {Ann. Phys.}\ }\textbf {\bibinfo
  {volume} {326}},\ \bibinfo {pages} {96} (\bibinfo {year} {2011})}\BibitemShut
  {NoStop}%
\bibitem [{\citenamefont {Biamonte}()}]{biamonte2020lectures}%
  \BibitemOpen
  \bibfield  {author} {\bibinfo {author} {\bibfnamefont {J.}~\bibnamefont
  {Biamonte}},\ }\href@noop {} {}\Eprint {https://arxiv.org/abs/1912.10049}
  {arXiv:1912.10049} \BibitemShut {NoStop}%
\bibitem [{\citenamefont {Silvi}\ \emph {et~al.}(2019)\citenamefont {Silvi},
  \citenamefont {Tschirsich}, \citenamefont {Gerster}, \citenamefont
  {Jünemann}, \citenamefont {Jaschke}, \citenamefont {Rizzi},\ and\
  \citenamefont {Montangero}}]{silvi2019the}%
  \BibitemOpen
  \bibfield  {author} {\bibinfo {author} {\bibfnamefont {P.}~\bibnamefont
  {Silvi}}, \bibinfo {author} {\bibfnamefont {F.}~\bibnamefont {Tschirsich}},
  \bibinfo {author} {\bibfnamefont {M.}~\bibnamefont {Gerster}}, \bibinfo
  {author} {\bibfnamefont {J.}~\bibnamefont {Jünemann}}, \bibinfo {author}
  {\bibfnamefont {D.}~\bibnamefont {Jaschke}}, \bibinfo {author} {\bibfnamefont
  {M.}~\bibnamefont {Rizzi}},\ and\ \bibinfo {author} {\bibfnamefont
  {S.}~\bibnamefont {Montangero}},\ }\href
  {https://doi.org/10.21468%2Fscipostphyslectnotes.8} {\bibfield  {journal}
  {\bibinfo  {journal} {{SciPost} Phys. Lect. Notes}\ } (\bibinfo {year}
  {2019})}\BibitemShut {NoStop}%
\bibitem [{\citenamefont {Orús}(2014)}]{Orus_2014}%
  \BibitemOpen
  \bibfield  {author} {\bibinfo {author} {\bibfnamefont {R.}~\bibnamefont
  {Orús}},\ }\href {https://doi.org/10.1016/j.aop.2014.06.013} {\bibfield
  {journal} {\bibinfo  {journal} {Annals of Physics}\ }\textbf {\bibinfo
  {volume} {349}},\ \bibinfo {pages} {117–158} (\bibinfo {year}
  {2014})}\BibitemShut {NoStop}%
\bibitem [{\citenamefont {Ranabhat}\ and\ \citenamefont
  {Collura}(2022)}]{Ranabhat_2022}%
  \BibitemOpen
  \bibfield  {author} {\bibinfo {author} {\bibfnamefont {N.}~\bibnamefont
  {Ranabhat}}\ and\ \bibinfo {author} {\bibfnamefont {M.}~\bibnamefont
  {Collura}},\ }\bibfield  {journal} {\bibinfo  {journal} {SciPost Physics}\
  }\textbf {\bibinfo {volume} {12}},\ \href
  {https://doi.org/10.21468/scipostphys.12.4.126}
  {10.21468/scipostphys.12.4.126} (\bibinfo {year} {2022})\BibitemShut
  {NoStop}%
\bibitem [{\citenamefont {Ranabhat}\ and\ \citenamefont
  {Collura}(2024)}]{Ranabhat_2024}%
  \BibitemOpen
  \bibfield  {author} {\bibinfo {author} {\bibfnamefont {N.}~\bibnamefont
  {Ranabhat}}\ and\ \bibinfo {author} {\bibfnamefont {M.}~\bibnamefont
  {Collura}},\ }\href {https://doi.org/10.21468/SciPostPhysCore.7.2.017}
  {\bibfield  {journal} {\bibinfo  {journal} {SciPost Phys. Core}\ }\textbf
  {\bibinfo {volume} {7}},\ \bibinfo {pages} {017} (\bibinfo {year}
  {2024})}\BibitemShut {NoStop}%
\bibitem [{\citenamefont {Verstraete}\ and\ \citenamefont
  {Cirac}(2004)}]{Verstraete_2004}%
  \BibitemOpen
  \bibfield  {author} {\bibinfo {author} {\bibfnamefont {F.}~\bibnamefont
  {Verstraete}}\ and\ \bibinfo {author} {\bibfnamefont {J.~I.}\ \bibnamefont
  {Cirac}},\ }\href {https://arxiv.org/abs/cond-mat/0407066} {} (\bibinfo
  {year} {2004}),\ \Eprint {https://arxiv.org/abs/cond-mat/0407066}
  {arXiv:cond-mat/0407066 [cond-mat.str-el]} \BibitemShut {NoStop}%
\bibitem [{\citenamefont {Jordan}\ \emph {et~al.}(2008)\citenamefont {Jordan},
  \citenamefont {Orus}, \citenamefont {Vidal}, \citenamefont {Verstraete},\
  and\ \citenamefont {Cirac}}]{Jordan_2008}%
  \BibitemOpen
  \bibfield  {author} {\bibinfo {author} {\bibfnamefont {J.}~\bibnamefont
  {Jordan}}, \bibinfo {author} {\bibfnamefont {R.}~\bibnamefont {Orus}},
  \bibinfo {author} {\bibfnamefont {G.}~\bibnamefont {Vidal}}, \bibinfo
  {author} {\bibfnamefont {F.}~\bibnamefont {Verstraete}},\ and\ \bibinfo
  {author} {\bibfnamefont {J.~I.}\ \bibnamefont {Cirac}},\ }\href
  {https://doi.org/10.1103/PhysRevLett.101.250602} {\bibfield  {journal}
  {\bibinfo  {journal} {Phys. Rev. Lett.}\ }\textbf {\bibinfo {volume} {101}},\
  \bibinfo {pages} {250602} (\bibinfo {year} {2008})}\BibitemShut {NoStop}%
\bibitem [{\citenamefont {Zaletel}\ and\ \citenamefont
  {Pollmann}(2020)}]{PhysRevLett.124.037201}%
  \BibitemOpen
  \bibfield  {author} {\bibinfo {author} {\bibfnamefont {M.~P.}\ \bibnamefont
  {Zaletel}}\ and\ \bibinfo {author} {\bibfnamefont {F.}~\bibnamefont
  {Pollmann}},\ }\href
  {https://link.aps.org/doi/10.1103/PhysRevLett.124.037201} {\bibfield
  {journal} {\bibinfo  {journal} {Phys. Rev. Lett.}\ }\textbf {\bibinfo
  {volume} {124}},\ \bibinfo {pages} {037201} (\bibinfo {year}
  {2020})}\BibitemShut {NoStop}%
\bibitem [{\citenamefont {Corboz}\ \emph {et~al.}(2011)\citenamefont {Corboz},
  \citenamefont {White}, \citenamefont {Vidal},\ and\ \citenamefont
  {Troyer}}]{Corboz_2011}%
  \BibitemOpen
  \bibfield  {author} {\bibinfo {author} {\bibfnamefont {P.}~\bibnamefont
  {Corboz}}, \bibinfo {author} {\bibfnamefont {S.~R.}\ \bibnamefont {White}},
  \bibinfo {author} {\bibfnamefont {G.}~\bibnamefont {Vidal}},\ and\ \bibinfo
  {author} {\bibfnamefont {M.}~\bibnamefont {Troyer}},\ }\href
  {https://doi.org/10.1103/PhysRevB.84.041108} {\bibfield  {journal} {\bibinfo
  {journal} {Phys. Rev. B}\ }\textbf {\bibinfo {volume} {84}},\ \bibinfo
  {pages} {041108} (\bibinfo {year} {2011})}\BibitemShut {NoStop}%
\bibitem [{\citenamefont {Piroli}\ \emph {et~al.}(2021)\citenamefont {Piroli},
  \citenamefont {Styliaris},\ and\ \citenamefont {Cirac}}]{piroli2021quantum}%
  \BibitemOpen
  \bibfield  {author} {\bibinfo {author} {\bibfnamefont {L.}~\bibnamefont
  {Piroli}}, \bibinfo {author} {\bibfnamefont {G.}~\bibnamefont {Styliaris}},\
  and\ \bibinfo {author} {\bibfnamefont {J.~I.}\ \bibnamefont {Cirac}},\ }\href
  {https://link.aps.org/doi/10.1103/PhysRevLett.127.220503} {\bibfield
  {journal} {\bibinfo  {journal} {Phys. Rev. Lett.}\ }\textbf {\bibinfo
  {volume} {127}},\ \bibinfo {pages} {220503} (\bibinfo {year}
  {2021})}\BibitemShut {NoStop}%
\bibitem [{\citenamefont {Malz}\ \emph {et~al.}(2024)\citenamefont {Malz},
  \citenamefont {Styliaris}, \citenamefont {Wei},\ and\ \citenamefont
  {Cirac}}]{malz2024preparation}%
  \BibitemOpen
  \bibfield  {author} {\bibinfo {author} {\bibfnamefont {D.}~\bibnamefont
  {Malz}}, \bibinfo {author} {\bibfnamefont {G.}~\bibnamefont {Styliaris}},
  \bibinfo {author} {\bibfnamefont {Z.-Y.}\ \bibnamefont {Wei}},\ and\ \bibinfo
  {author} {\bibfnamefont {J.~I.}\ \bibnamefont {Cirac}},\ }\href
  {https://link.aps.org/doi/10.1103/PhysRevLett.132.040404} {\bibfield
  {journal} {\bibinfo  {journal} {Phys. Rev. Lett.}\ }\textbf {\bibinfo
  {volume} {132}},\ \bibinfo {pages} {040404} (\bibinfo {year}
  {2024})}\BibitemShut {NoStop}%
\bibitem [{\citenamefont {Smith}\ \emph {et~al.}()\citenamefont {Smith},
  \citenamefont {Khan}, \citenamefont {Clark}, \citenamefont {Girvin},\ and\
  \citenamefont {Wei}}]{smith2024kevin}%
  \BibitemOpen
  \bibfield  {author} {\bibinfo {author} {\bibfnamefont {K.~C.}\ \bibnamefont
  {Smith}}, \bibinfo {author} {\bibfnamefont {A.}~\bibnamefont {Khan}},
  \bibinfo {author} {\bibfnamefont {B.~K.}\ \bibnamefont {Clark}}, \bibinfo
  {author} {\bibfnamefont {S.~M.}\ \bibnamefont {Girvin}},\ and\ \bibinfo
  {author} {\bibfnamefont {T.-C.}\ \bibnamefont {Wei}},\ }\href@noop {}
  {}\Eprint {https://arxiv.org/abs/2404.16083} {arXiv:2404.16083} \BibitemShut
  {NoStop}%
\bibitem [{\citenamefont {Stephen}\ and\ \citenamefont
  {Hart}()}]{david2024preparing}%
  \BibitemOpen
  \bibfield  {author} {\bibinfo {author} {\bibfnamefont {D.~T.}\ \bibnamefont
  {Stephen}}\ and\ \bibinfo {author} {\bibfnamefont {O.}~\bibnamefont {Hart}},\
  }\href@noop {} {}\Eprint {https://arxiv.org/abs/2404.16360}
  {arXiv:2404.16360} \BibitemShut {NoStop}%
\bibitem [{\citenamefont {Zhang}\ \emph {et~al.}()\citenamefont {Zhang},
  \citenamefont {Gopalakrishnan},\ and\ \citenamefont
  {Styliaris}}]{zhang2024yifan}%
  \BibitemOpen
  \bibfield  {author} {\bibinfo {author} {\bibfnamefont {Y.}~\bibnamefont
  {Zhang}}, \bibinfo {author} {\bibfnamefont {S.}~\bibnamefont
  {Gopalakrishnan}},\ and\ \bibinfo {author} {\bibfnamefont {G.}~\bibnamefont
  {Styliaris}},\ }\href@noop {} {}\Eprint {https://arxiv.org/abs/2405.09615}
  {arXiv:2405.09615} \BibitemShut {NoStop}%
\bibitem [{\citenamefont {Garnerone}\ \emph
  {et~al.}(2010{\natexlab{a}})\citenamefont {Garnerone}, \citenamefont
  {de~Oliveira},\ and\ \citenamefont {Zanardi}}]{garnerone2010typicality}%
  \BibitemOpen
  \bibfield  {author} {\bibinfo {author} {\bibfnamefont {S.}~\bibnamefont
  {Garnerone}}, \bibinfo {author} {\bibfnamefont {T.~R.}\ \bibnamefont
  {de~Oliveira}},\ and\ \bibinfo {author} {\bibfnamefont {P.}~\bibnamefont
  {Zanardi}},\ }\href {http://dx.doi.org/10.1103/PhysRevA.81.032336} {\bibfield
   {journal} {\bibinfo  {journal} {Phys. Rev. A}\ }\textbf {\bibinfo {volume}
  {81}},\ \bibinfo {pages} {032336} (\bibinfo {year}
  {2010}{\natexlab{a}})}\BibitemShut {NoStop}%
\bibitem [{\citenamefont {Garnerone}\ \emph
  {et~al.}(2010{\natexlab{b}})\citenamefont {Garnerone}, \citenamefont
  {de~Oliveira}, \citenamefont {Haas},\ and\ \citenamefont
  {Zanardi}}]{garnerone2010statistical}%
  \BibitemOpen
  \bibfield  {author} {\bibinfo {author} {\bibfnamefont {S.}~\bibnamefont
  {Garnerone}}, \bibinfo {author} {\bibfnamefont {T.~R.}\ \bibnamefont
  {de~Oliveira}}, \bibinfo {author} {\bibfnamefont {S.}~\bibnamefont {Haas}},\
  and\ \bibinfo {author} {\bibfnamefont {P.}~\bibnamefont {Zanardi}},\ }\href
  {http://dx.doi.org/10.1103/PhysRevA.82.052312} {\bibfield  {journal}
  {\bibinfo  {journal} {Phys. Rev. A}\ }\textbf {\bibinfo {volume} {82}},\
  \bibinfo {pages} {052312} (\bibinfo {year} {2010}{\natexlab{b}})}\BibitemShut
  {NoStop}%
\bibitem [{\citenamefont {Lancien}\ and\ \citenamefont
  {Pérez-García}(2021)}]{lancien2021correlation}%
  \BibitemOpen
  \bibfield  {author} {\bibinfo {author} {\bibfnamefont {C.}~\bibnamefont
  {Lancien}}\ and\ \bibinfo {author} {\bibfnamefont {D.}~\bibnamefont
  {Pérez-García}},\ }\href {http://dx.doi.org/10.1007/s00023-021-01087-4}
  {\bibfield  {journal} {\bibinfo  {journal} {Annales Henri Poincaré}\
  }\textbf {\bibinfo {volume} {23}},\ \bibinfo {pages} {141–222} (\bibinfo
  {year} {2021})}\BibitemShut {NoStop}%
\bibitem [{\citenamefont {Haferkamp}\ \emph {et~al.}(2021)\citenamefont
  {Haferkamp}, \citenamefont {Bertoni}, \citenamefont {Roth},\ and\
  \citenamefont {Eisert}}]{haferkamp2021emergent}%
  \BibitemOpen
  \bibfield  {author} {\bibinfo {author} {\bibfnamefont {J.}~\bibnamefont
  {Haferkamp}}, \bibinfo {author} {\bibfnamefont {C.}~\bibnamefont {Bertoni}},
  \bibinfo {author} {\bibfnamefont {I.}~\bibnamefont {Roth}},\ and\ \bibinfo
  {author} {\bibfnamefont {J.}~\bibnamefont {Eisert}},\ }\href
  {http://dx.doi.org/10.1103/PRXQuantum.2.040308} {\bibfield  {journal}
  {\bibinfo  {journal} {PRX Quantum}\ }\textbf {\bibinfo {volume} {2}}
  (\bibinfo {year} {2021})}\BibitemShut {NoStop}%
\bibitem [{\citenamefont {Lami}\ \emph {et~al.}()\citenamefont {Lami},
  \citenamefont {Haug},\ and\ \citenamefont {Nardis}}]{lami2024quantum}%
  \BibitemOpen
  \bibfield  {author} {\bibinfo {author} {\bibfnamefont {G.}~\bibnamefont
  {Lami}}, \bibinfo {author} {\bibfnamefont {T.}~\bibnamefont {Haug}},\ and\
  \bibinfo {author} {\bibfnamefont {J.~D.}\ \bibnamefont {Nardis}},\
  }\href@noop {} {}\Eprint {https://arxiv.org/abs/2404.18751}
  {arXiv:2404.18751} \BibitemShut {NoStop}%
\bibitem [{\citenamefont {Haag}\ \emph {et~al.}(2023)\citenamefont {Haag},
  \citenamefont {Baccari},\ and\ \citenamefont {Styliaris}}]{haag2023typical}%
  \BibitemOpen
  \bibfield  {author} {\bibinfo {author} {\bibfnamefont {D.}~\bibnamefont
  {Haag}}, \bibinfo {author} {\bibfnamefont {F.}~\bibnamefont {Baccari}},\ and\
  \bibinfo {author} {\bibfnamefont {G.}~\bibnamefont {Styliaris}},\ }\href
  {http://dx.doi.org/10.1103/PRXQuantum.4.030330} {\bibfield  {journal}
  {\bibinfo  {journal} {PRX Quantum}\ }\textbf {\bibinfo {volume} {4}},\
  \bibinfo {pages} {030330} (\bibinfo {year} {2023})}\BibitemShut {NoStop}%
\bibitem [{\citenamefont {Metger}\ \emph {et~al.}(2024)\citenamefont {Metger},
  \citenamefont {Poremba}, \citenamefont {Sinha},\ and\ \citenamefont
  {Yuen}}]{Yuen2024}%
  \BibitemOpen
  \bibfield  {author} {\bibinfo {author} {\bibfnamefont {T.}~\bibnamefont
  {Metger}}, \bibinfo {author} {\bibfnamefont {A.}~\bibnamefont {Poremba}},
  \bibinfo {author} {\bibfnamefont {M.}~\bibnamefont {Sinha}},\ and\ \bibinfo
  {author} {\bibfnamefont {H.}~\bibnamefont {Yuen}},\ }\href@noop {} {\bibinfo
  {title} {Simple constructions of linear-depth t-designs and pseudorandom
  unitaries}} (\bibinfo {year} {2024}),\ \Eprint
  {https://arxiv.org/abs/arXiv:2404.12647} {arXiv:2404.12647} \BibitemShut
  {NoStop}%
\bibitem [{\citenamefont {Porter}\ and\ \citenamefont
  {Thomas}(1956)}]{porter1956}%
  \BibitemOpen
  \bibfield  {author} {\bibinfo {author} {\bibfnamefont {C.~E.}\ \bibnamefont
  {Porter}}\ and\ \bibinfo {author} {\bibfnamefont {R.~G.}\ \bibnamefont
  {Thomas}},\ }\href {https://doi.org/10.1103/PhysRev.104.483} {\bibfield
  {journal} {\bibinfo  {journal} {Phys. Rev.}\ }\textbf {\bibinfo {volume}
  {104}},\ \bibinfo {pages} {483} (\bibinfo {year} {1956})}\BibitemShut
  {NoStop}%
\bibitem [{\citenamefont {Haake}(2001)}]{haake2001}%
  \BibitemOpen
  \bibfield  {author} {\bibinfo {author} {\bibfnamefont {F.}~\bibnamefont
  {Haake}},\ }\href@noop {} {\emph {\bibinfo {title} {Quantum Signatures of
  Chaos}}}\ (\bibinfo  {publisher} {Springer, Heidelberg, Germany},\ \bibinfo
  {year} {2001})\BibitemShut {NoStop}%
\bibitem [{\citenamefont {Mullane}()}]{Mullane2020}%
  \BibitemOpen
  \bibfield  {author} {\bibinfo {author} {\bibfnamefont {S.}~\bibnamefont
  {Mullane}},\ }\href@noop {} {}\Eprint {https://arxiv.org/abs/2007.07872}
  {arXiv:2007.07872} \BibitemShut {NoStop}%
\bibitem [{\citenamefont {Turkeshi}\ \emph {et~al.}()\citenamefont {Turkeshi},
  \citenamefont {Dymarsky},\ and\ \citenamefont
  {Sierant}}]{turkeshi2023paulispectrummagictypical}%
  \BibitemOpen
  \bibfield  {author} {\bibinfo {author} {\bibfnamefont {X.}~\bibnamefont
  {Turkeshi}}, \bibinfo {author} {\bibfnamefont {A.}~\bibnamefont {Dymarsky}},\
  and\ \bibinfo {author} {\bibfnamefont {P.}~\bibnamefont {Sierant}},\
  }\href@noop {} {}\Eprint {https://arxiv.org/abs/2312.11631}
  {arXiv:2312.11631} \BibitemShut {NoStop}%
\bibitem [{\citenamefont {Turkeshi}\ \emph {et~al.}(2023)\citenamefont
  {Turkeshi}, \citenamefont {Schir\`o},\ and\ \citenamefont
  {Sierant}}]{turkeshi2023measuring}%
  \BibitemOpen
  \bibfield  {author} {\bibinfo {author} {\bibfnamefont {X.}~\bibnamefont
  {Turkeshi}}, \bibinfo {author} {\bibfnamefont {M.}~\bibnamefont {Schir\`o}},\
  and\ \bibinfo {author} {\bibfnamefont {P.}~\bibnamefont {Sierant}},\ }\href
  {https://link.aps.org/doi/10.1103/PhysRevA.108.042408} {\bibfield  {journal}
  {\bibinfo  {journal} {Phys. Rev. A}\ }\textbf {\bibinfo {volume} {108}},\
  \bibinfo {pages} {042408} (\bibinfo {year} {2023})}\BibitemShut {NoStop}%
\bibitem [{\citenamefont {Choi}(1975)}]{CHOI1975285}%
  \BibitemOpen
  \bibfield  {author} {\bibinfo {author} {\bibfnamefont {M.-D.}\ \bibnamefont
  {Choi}},\ }\href@noop {} {\bibfield  {journal} {\bibinfo  {journal} {Linear
  Algebra and its Applications}\ }\textbf {\bibinfo {volume} {10}},\ \bibinfo
  {pages} {285} (\bibinfo {year} {1975})}\BibitemShut {NoStop}%
\bibitem [{\citenamefont {Köstenberger}()}]{Kostenberger_2021}%
  \BibitemOpen
  \bibfield  {author} {\bibinfo {author} {\bibfnamefont {G.}~\bibnamefont
  {Köstenberger}},\ }\href@noop {} {}\Eprint
  {https://arxiv.org/abs/2101.00921} {arXiv:2101.00921} \BibitemShut {NoStop}%
\bibitem [{\citenamefont {Chan}\ \emph {et~al.}(2022)\citenamefont {Chan},
  \citenamefont {Shivam}, \citenamefont {Huse},\ and\ \citenamefont
  {De~Luca}}]{Chan2022}%
  \BibitemOpen
  \bibfield  {author} {\bibinfo {author} {\bibfnamefont {A.}~\bibnamefont
  {Chan}}, \bibinfo {author} {\bibfnamefont {S.}~\bibnamefont {Shivam}},
  \bibinfo {author} {\bibfnamefont {D.~A.}\ \bibnamefont {Huse}},\ and\
  \bibinfo {author} {\bibfnamefont {A.}~\bibnamefont {De~Luca}},\ }\href
  {http://dx.doi.org/10.1038/s41467-022-34318-1} {\bibfield  {journal}
  {\bibinfo  {journal} {Nature Comm.}\ }\textbf {\bibinfo {volume} {13}},\
  \bibinfo {pages} {7484} (\bibinfo {year} {2022})}\BibitemShut {NoStop}%
\bibitem [{\citenamefont {Loio}\ \emph {et~al.}(2024)\citenamefont {Loio},
  \citenamefont {Cecile}, \citenamefont {Gopalakrishnan}, \citenamefont
  {Lami},\ and\ \citenamefont {Nardis}}]{Loio2024Toappear}%
  \BibitemOpen
  \bibfield  {author} {\bibinfo {author} {\bibfnamefont {H.}~\bibnamefont
  {Loio}}, \bibinfo {author} {\bibfnamefont {G.}~\bibnamefont {Cecile}},
  \bibinfo {author} {\bibfnamefont {S.}~\bibnamefont {Gopalakrishnan}},
  \bibinfo {author} {\bibfnamefont {G.}~\bibnamefont {Lami}},\ and\ \bibinfo
  {author} {\bibfnamefont {J.~D.}\ \bibnamefont {Nardis}},\ }\href@noop {}
  {\bibfield  {journal} {\bibinfo  {journal} {To appear}\ } (\bibinfo {year}
  {2024})}\BibitemShut {NoStop}%
\bibitem [{\citenamefont {Stoudenmire}\ and\ \citenamefont
  {White}(2010)}]{Stoudenmire_2010}%
  \BibitemOpen
  \bibfield  {author} {\bibinfo {author} {\bibfnamefont {E.~M.}\ \bibnamefont
  {Stoudenmire}}\ and\ \bibinfo {author} {\bibfnamefont {S.~R.}\ \bibnamefont
  {White}},\ }\href {https://doi.org/10.1088/1367-2630/12/5/055026} {\bibfield
  {journal} {\bibinfo  {journal} {New J. Phys.}\ }\textbf {\bibinfo {volume}
  {12}},\ \bibinfo {pages} {055026} (\bibinfo {year} {2010})}\BibitemShut
  {NoStop}%
\bibitem [{\citenamefont {Lami}\ and\ \citenamefont
  {Collura}(2023)}]{Lami_2023_2}%
  \BibitemOpen
  \bibfield  {author} {\bibinfo {author} {\bibfnamefont {G.}~\bibnamefont
  {Lami}}\ and\ \bibinfo {author} {\bibfnamefont {M.}~\bibnamefont {Collura}},\
  }\href {https://link.aps.org/doi/10.1103/PhysRevLett.131.180401} {\bibfield
  {journal} {\bibinfo  {journal} {Phys. Rev. Lett.}\ }\textbf {\bibinfo
  {volume} {131}},\ \bibinfo {pages} {180401} (\bibinfo {year}
  {2023})}\BibitemShut {NoStop}%
\bibitem [{\citenamefont {Lami}\ and\ \citenamefont
  {Collura}(2024)}]{Lami_2024}%
  \BibitemOpen
  \bibfield  {author} {\bibinfo {author} {\bibfnamefont {G.}~\bibnamefont
  {Lami}}\ and\ \bibinfo {author} {\bibfnamefont {M.}~\bibnamefont {Collura}},\
  }\href {https://link.aps.org/doi/10.1103/PhysRevLett.133.010602} {\bibfield
  {journal} {\bibinfo  {journal} {Phys. Rev. Lett.}\ }\textbf {\bibinfo
  {volume} {133}},\ \bibinfo {pages} {010602} (\bibinfo {year}
  {2024})}\BibitemShut {NoStop}%
\bibitem [{\citenamefont {Lami}\ \emph {et~al.}(2023)\citenamefont {Lami},
  \citenamefont {Torta}, \citenamefont {Santoro},\ and\ \citenamefont
  {Collura}}]{Lami_2023_1}%
  \BibitemOpen
  \bibfield  {author} {\bibinfo {author} {\bibfnamefont {G.}~\bibnamefont
  {Lami}}, \bibinfo {author} {\bibfnamefont {P.}~\bibnamefont {Torta}},
  \bibinfo {author} {\bibfnamefont {G.~E.}\ \bibnamefont {Santoro}},\ and\
  \bibinfo {author} {\bibfnamefont {M.}~\bibnamefont {Collura}},\ }\href
  {http://dx.doi.org/10.21468/SciPostPhys.14.5.117} {\bibfield  {journal}
  {\bibinfo  {journal} {SciPost Phys.}\ }\textbf {\bibinfo {volume} {14}}
  (\bibinfo {year} {2023})}\BibitemShut {NoStop}%
\bibitem [{\citenamefont {Nahum}\ \emph {et~al.}(2017)\citenamefont {Nahum},
  \citenamefont {Ruhman}, \citenamefont {Vijay},\ and\ \citenamefont
  {Haah}}]{nahum2017quantum}%
  \BibitemOpen
  \bibfield  {author} {\bibinfo {author} {\bibfnamefont {A.}~\bibnamefont
  {Nahum}}, \bibinfo {author} {\bibfnamefont {J.}~\bibnamefont {Ruhman}},
  \bibinfo {author} {\bibfnamefont {S.}~\bibnamefont {Vijay}},\ and\ \bibinfo
  {author} {\bibfnamefont {J.}~\bibnamefont {Haah}},\ }\href
  {https://doi.org/10.1103/PhysRevX.7.031016} {\bibfield  {journal} {\bibinfo
  {journal} {Phys. Rev. X}\ }\textbf {\bibinfo {volume} {7}},\ \bibinfo {pages}
  {031016} (\bibinfo {year} {2017})}\BibitemShut {NoStop}%
\bibitem [{\citenamefont {Aaronson}\ \emph {et~al.}()\citenamefont {Aaronson},
  \citenamefont {Bouland}, \citenamefont {Fefferman}, \citenamefont {Ghosh},
  \citenamefont {Vazirani}, \citenamefont {Zhang},\ and\ \citenamefont
  {Zhou}}]{AaronsonPseudo}%
  \BibitemOpen
  \bibfield  {author} {\bibinfo {author} {\bibfnamefont {S.}~\bibnamefont
  {Aaronson}}, \bibinfo {author} {\bibfnamefont {A.}~\bibnamefont {Bouland}},
  \bibinfo {author} {\bibfnamefont {B.}~\bibnamefont {Fefferman}}, \bibinfo
  {author} {\bibfnamefont {S.}~\bibnamefont {Ghosh}}, \bibinfo {author}
  {\bibfnamefont {U.}~\bibnamefont {Vazirani}}, \bibinfo {author}
  {\bibfnamefont {C.}~\bibnamefont {Zhang}},\ and\ \bibinfo {author}
  {\bibfnamefont {Z.}~\bibnamefont {Zhou}},\ }\href@noop {} {}\Eprint
  {https://arxiv.org/abs/2211.00747} {arXiv:2211.00747} \BibitemShut {NoStop}%
\bibitem [{\citenamefont {Feng}\ and\ \citenamefont
  {Ippoliti}()}]{IppolitiPseudo}%
  \BibitemOpen
  \bibfield  {author} {\bibinfo {author} {\bibfnamefont {X.}~\bibnamefont
  {Feng}}\ and\ \bibinfo {author} {\bibfnamefont {M.}~\bibnamefont
  {Ippoliti}},\ }\href@noop {} {}\Eprint {https://arxiv.org/abs/2403.09619}
  {arXiv:2403.09619} \BibitemShut {NoStop}%
\bibitem [{\citenamefont {Grewal}\ \emph {et~al.}(2024)\citenamefont {Grewal},
  \citenamefont {Iyer}, \citenamefont {Kretschmer},\ and\ \citenamefont
  {Liang}}]{GrewalPseudo}%
  \BibitemOpen
  \bibfield  {author} {\bibinfo {author} {\bibfnamefont {S.}~\bibnamefont
  {Grewal}}, \bibinfo {author} {\bibfnamefont {V.}~\bibnamefont {Iyer}},
  \bibinfo {author} {\bibfnamefont {W.}~\bibnamefont {Kretschmer}},\ and\
  \bibinfo {author} {\bibfnamefont {D.}~\bibnamefont {Liang}},\ }\href
  {https://arxiv.org/abs/2404.00126} {\bibinfo {title} {{Pseudoentanglement
  Ain't Cheap}}} (\bibinfo {year} {2024}),\ \Eprint
  {https://arxiv.org/abs/2404.00126} {arXiv:2404.00126} \BibitemShut {NoStop}%
\bibitem [{\citenamefont {Gu}\ \emph {et~al.}(2024)\citenamefont {Gu},
  \citenamefont {Leone}, \citenamefont {Ghosh}, \citenamefont {Eisert},
  \citenamefont {Yelin},\ and\ \citenamefont {Quek}}]{gu2024pseudomagic}%
  \BibitemOpen
  \bibfield  {author} {\bibinfo {author} {\bibfnamefont {A.}~\bibnamefont
  {Gu}}, \bibinfo {author} {\bibfnamefont {L.}~\bibnamefont {Leone}}, \bibinfo
  {author} {\bibfnamefont {S.}~\bibnamefont {Ghosh}}, \bibinfo {author}
  {\bibfnamefont {J.}~\bibnamefont {Eisert}}, \bibinfo {author} {\bibfnamefont
  {S.~F.}\ \bibnamefont {Yelin}},\ and\ \bibinfo {author} {\bibfnamefont
  {Y.}~\bibnamefont {Quek}},\ }\href
  {https://link.aps.org/doi/10.1103/PhysRevLett.132.210602} {\bibfield
  {journal} {\bibinfo  {journal} {Phys. Rev. Lett.}\ }\textbf {\bibinfo
  {volume} {132}},\ \bibinfo {pages} {210602} (\bibinfo {year}
  {2024})}\BibitemShut {NoStop}%
\bibitem [{\citenamefont {Cheng}\ \emph {et~al.}({\natexlab{b}})\citenamefont
  {Cheng}, \citenamefont {Feng},\ and\ \citenamefont {Ippoliti}}]{ippo}%
  \BibitemOpen
  \bibfield  {author} {\bibinfo {author} {\bibfnamefont {Z.}~\bibnamefont
  {Cheng}}, \bibinfo {author} {\bibfnamefont {X.}~\bibnamefont {Feng}},\ and\
  \bibinfo {author} {\bibfnamefont {M.}~\bibnamefont {Ippoliti}},\ }\href@noop
  {} {\bibinfo {title} {To appear.}} ({\natexlab{b}})\BibitemShut {NoStop}%
\bibitem [{\citenamefont {Paviglianiti}\ \emph {et~al.}()\citenamefont
  {Paviglianiti}, \citenamefont {Lami}, \citenamefont {Collura},\ and\
  \citenamefont {Silva}}]{Paviglianiti_2024}%
  \BibitemOpen
  \bibfield  {author} {\bibinfo {author} {\bibfnamefont {A.}~\bibnamefont
  {Paviglianiti}}, \bibinfo {author} {\bibfnamefont {G.}~\bibnamefont {Lami}},
  \bibinfo {author} {\bibfnamefont {M.}~\bibnamefont {Collura}},\ and\ \bibinfo
  {author} {\bibfnamefont {A.}~\bibnamefont {Silva}},\ }\href@noop {} {}\Eprint
  {https://arxiv.org/abs/2405.06054} {arXiv:2405.06054} \BibitemShut {NoStop}%
\bibitem [{\citenamefont {Mello}\ \emph {et~al.}(2024)\citenamefont {Mello},
  \citenamefont {Santini}, \citenamefont {Lami}, \citenamefont {Nardis},\ and\
  \citenamefont {Collura}}]{MelloSantini2024}%
  \BibitemOpen
  \bibfield  {author} {\bibinfo {author} {\bibfnamefont {A.~F.}\ \bibnamefont
  {Mello}}, \bibinfo {author} {\bibfnamefont {A.}~\bibnamefont {Santini}},
  \bibinfo {author} {\bibfnamefont {G.}~\bibnamefont {Lami}}, \bibinfo {author}
  {\bibfnamefont {J.~D.}\ \bibnamefont {Nardis}},\ and\ \bibinfo {author}
  {\bibfnamefont {M.}~\bibnamefont {Collura}},\ }\href@noop {} {} (\bibinfo
  {year} {2024}),\ \Eprint {https://arxiv.org/abs/2407.01692}
  {arXiv:2407.01692} \BibitemShut {NoStop}%
\bibitem [{\citenamefont {Xiangjian}\ \emph
  {et~al.}({\natexlab{a}})\citenamefont {Xiangjian}, \citenamefont {Jiale},\
  and\ \citenamefont {Qin}}]{Qian2024}%
  \BibitemOpen
  \bibfield  {author} {\bibinfo {author} {\bibfnamefont {Q.}~\bibnamefont
  {Xiangjian}}, \bibinfo {author} {\bibfnamefont {H.}~\bibnamefont {Jiale}},\
  and\ \bibinfo {author} {\bibfnamefont {M.}~\bibnamefont {Qin}},\ }\href@noop
  {} {} ({\natexlab{a}}),\ \Eprint {https://arxiv.org/abs/2407.03202}
  {arXiv:2407.03202} \BibitemShut {NoStop}%
\bibitem [{\citenamefont {Xiangjian}\ \emph
  {et~al.}({\natexlab{b}})\citenamefont {Xiangjian}, \citenamefont {Jiale},\
  and\ \citenamefont {Mingpu}}]{Qian20242}%
  \BibitemOpen
  \bibfield  {author} {\bibinfo {author} {\bibfnamefont {Q.}~\bibnamefont
  {Xiangjian}}, \bibinfo {author} {\bibfnamefont {H.}~\bibnamefont {Jiale}},\
  and\ \bibinfo {author} {\bibfnamefont {Q.}~\bibnamefont {Mingpu}},\
  }\href@noop {} {} ({\natexlab{b}}),\ \Eprint
  {https://arxiv.org/abs/2405.09217} {arXiv:2405.09217} \BibitemShut {NoStop}%
\bibitem [{\citenamefont {Dowling}\ \emph {et~al.}()\citenamefont {Dowling},
  \citenamefont {Kos},\ and\ \citenamefont
  {Turkeshi}}]{dowling2024magicheisenbergpicture}%
  \BibitemOpen
  \bibfield  {author} {\bibinfo {author} {\bibfnamefont {N.}~\bibnamefont
  {Dowling}}, \bibinfo {author} {\bibfnamefont {P.}~\bibnamefont {Kos}},\ and\
  \bibinfo {author} {\bibfnamefont {X.}~\bibnamefont {Turkeshi}},\ }\href@noop
  {} {}\Eprint {https://arxiv.org/abs/2408.16047} {arXiv:2408.16047}
  \BibitemShut {NoStop}%
\bibitem [{\citenamefont {Iwaki}\ \emph {et~al.}(2021)\citenamefont {Iwaki},
  \citenamefont {Shimizu},\ and\ \citenamefont
  {Hotta}}]{PhysRevResearch.3.L022015}%
  \BibitemOpen
  \bibfield  {author} {\bibinfo {author} {\bibfnamefont {A.}~\bibnamefont
  {Iwaki}}, \bibinfo {author} {\bibfnamefont {A.}~\bibnamefont {Shimizu}},\
  and\ \bibinfo {author} {\bibfnamefont {C.}~\bibnamefont {Hotta}},\ }\href
  {https://link.aps.org/doi/10.1103/PhysRevResearch.3.L022015} {\bibfield
  {journal} {\bibinfo  {journal} {Phys. Rev. Res.}\ }\textbf {\bibinfo {volume}
  {3}},\ \bibinfo {pages} {L022015} (\bibinfo {year} {2021})}\BibitemShut
  {NoStop}%
\bibitem [{\citenamefont {Fr\'{\i}as-P\'erez}\ \emph
  {et~al.}(2024)\citenamefont {Fr\'{\i}as-P\'erez}, \citenamefont
  {Tagliacozzo},\ and\ \citenamefont {Ba\~nuls}}]{PhysRevLett.132.100402}%
  \BibitemOpen
  \bibfield  {author} {\bibinfo {author} {\bibfnamefont {M.}~\bibnamefont
  {Fr\'{\i}as-P\'erez}}, \bibinfo {author} {\bibfnamefont {L.}~\bibnamefont
  {Tagliacozzo}},\ and\ \bibinfo {author} {\bibfnamefont {M.~C.}\ \bibnamefont
  {Ba\~nuls}},\ }\href
  {https://link.aps.org/doi/10.1103/PhysRevLett.132.100402} {\bibfield
  {journal} {\bibinfo  {journal} {Phys. Rev. Lett.}\ }\textbf {\bibinfo
  {volume} {132}},\ \bibinfo {pages} {100402} (\bibinfo {year}
  {2024})}\BibitemShut {NoStop}%
\bibitem [{\citenamefont {Rakovszky}\ \emph {et~al.}(2022)\citenamefont
  {Rakovszky}, \citenamefont {von Keyserlingk},\ and\ \citenamefont
  {Pollmann}}]{PhysRevB.105.075131}%
  \BibitemOpen
  \bibfield  {author} {\bibinfo {author} {\bibfnamefont {T.}~\bibnamefont
  {Rakovszky}}, \bibinfo {author} {\bibfnamefont {C.~W.}\ \bibnamefont {von
  Keyserlingk}},\ and\ \bibinfo {author} {\bibfnamefont {F.}~\bibnamefont
  {Pollmann}},\ }\href {https://link.aps.org/doi/10.1103/PhysRevB.105.075131}
  {\bibfield  {journal} {\bibinfo  {journal} {Phys. Rev. B}\ }\textbf {\bibinfo
  {volume} {105}},\ \bibinfo {pages} {075131} (\bibinfo {year}
  {2022})}\BibitemShut {NoStop}%
\bibitem [{\citenamefont {Khemani}\ \emph {et~al.}(2018)\citenamefont
  {Khemani}, \citenamefont {Vishwanath},\ and\ \citenamefont
  {Huse}}]{PhysRevX.8.031057}%
  \BibitemOpen
  \bibfield  {author} {\bibinfo {author} {\bibfnamefont {V.}~\bibnamefont
  {Khemani}}, \bibinfo {author} {\bibfnamefont {A.}~\bibnamefont
  {Vishwanath}},\ and\ \bibinfo {author} {\bibfnamefont {D.~A.}\ \bibnamefont
  {Huse}},\ }\href {https://link.aps.org/doi/10.1103/PhysRevX.8.031057}
  {\bibfield  {journal} {\bibinfo  {journal} {Phys. Rev. X}\ }\textbf {\bibinfo
  {volume} {8}},\ \bibinfo {pages} {031057} (\bibinfo {year}
  {2018})}\BibitemShut {NoStop}%
\bibitem [{\citenamefont {Rakovszky}\ \emph {et~al.}(2018)\citenamefont
  {Rakovszky}, \citenamefont {Pollmann},\ and\ \citenamefont {von
  Keyserlingk}}]{PhysRevX.8.031058}%
  \BibitemOpen
  \bibfield  {author} {\bibinfo {author} {\bibfnamefont {T.}~\bibnamefont
  {Rakovszky}}, \bibinfo {author} {\bibfnamefont {F.}~\bibnamefont
  {Pollmann}},\ and\ \bibinfo {author} {\bibfnamefont {C.~W.}\ \bibnamefont
  {von Keyserlingk}},\ }\href
  {https://link.aps.org/doi/10.1103/PhysRevX.8.031058} {\bibfield  {journal}
  {\bibinfo  {journal} {Phys. Rev. X}\ }\textbf {\bibinfo {volume} {8}},\
  \bibinfo {pages} {031058} (\bibinfo {year} {2018})}\BibitemShut {NoStop}%
\bibitem [{\citenamefont {Ba\~nuls}\ \emph {et~al.}(2017)\citenamefont
  {Ba\~nuls}, \citenamefont {Cichy}, \citenamefont {Cirac}, \citenamefont
  {Jansen},\ and\ \citenamefont {K\"uhn}}]{PhysRevX.7.041046}%
  \BibitemOpen
  \bibfield  {author} {\bibinfo {author} {\bibfnamefont {M.~C.}\ \bibnamefont
  {Ba\~nuls}}, \bibinfo {author} {\bibfnamefont {K.}~\bibnamefont {Cichy}},
  \bibinfo {author} {\bibfnamefont {J.~I.}\ \bibnamefont {Cirac}}, \bibinfo
  {author} {\bibfnamefont {K.}~\bibnamefont {Jansen}},\ and\ \bibinfo {author}
  {\bibfnamefont {S.}~\bibnamefont {K\"uhn}},\ }\href
  {https://link.aps.org/doi/10.1103/PhysRevX.7.041046} {\bibfield  {journal}
  {\bibinfo  {journal} {Phys. Rev. X}\ }\textbf {\bibinfo {volume} {7}},\
  \bibinfo {pages} {041046} (\bibinfo {year} {2017})}\BibitemShut {NoStop}%
\bibitem [{Note1()}]{Note1}%
  \BibitemOpen
  \bibinfo {note} {In fact, $f_\lambda = \chi _\lambda (())$ the character of
  the identity permutation, and $\DOTSB \tsum \slimits@ _{\lambda \vdash k}
  f_\lambda ^2 = k!$, consistently.}\BibitemShut {Stop}%
\end{thebibliography}%
\clearpage
\setcounter{section}{0}
\setcounter{secnumdepth}{2}

\begin{center}
    \textbf{End Matter}
\end{center}

\section{Random product states}

\begin{figure}[h!]
\centering
\includegraphics[width=0.85\linewidth]{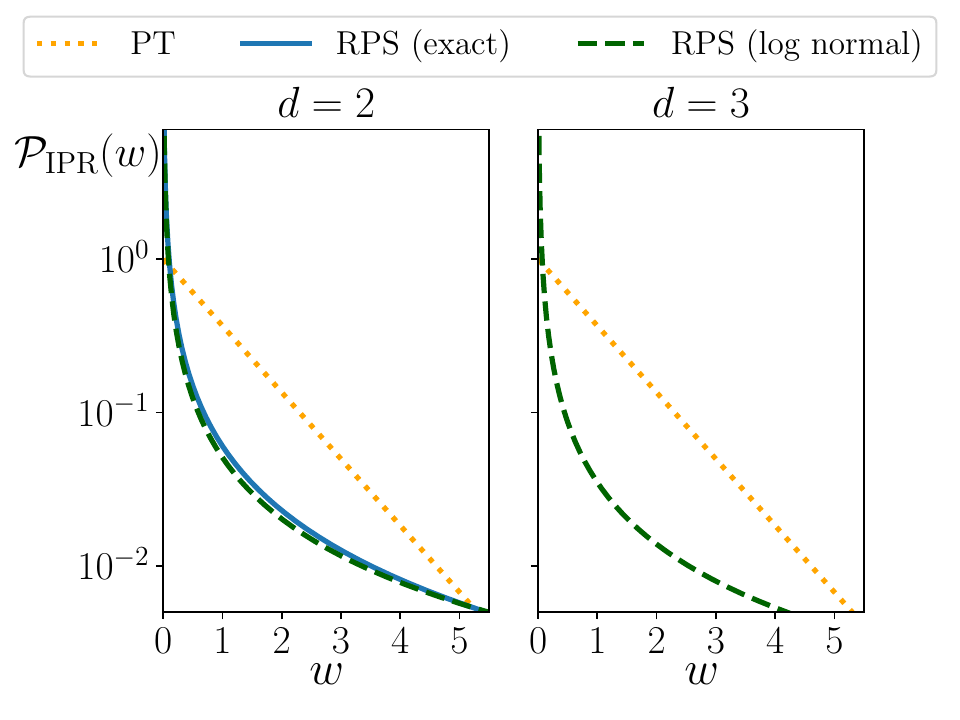}
\caption{Probability distribution of overlaps over the computational basis $\mathcal{P}_{\text{IPR}}(w)$ for Haar and RMPS. The latter is obtained with the exact formula (only for $d=2$), or with the log normal approximation. We set $N=15$ and $d=2,3$.} 
\label{fig:pwrps}
\end{figure}

In this section, we consider the ensemble $\mathcal{E}$ of Random Product States (RPS), which can be viewed as RMPS with a bond dimension of $\chi=1$. RPS are defined as $\ket{\psi} = \bigotimes_{i=1}^N \ket{\phi_i}$, with $\ket{\phi_i} \sim \Haar$. The distribution of overlaps $w_i = d |\braket{\phi_i | \phi_i'}|^{2}$ for a single qudit is given by the PT: $\mathcal{P}(w_i) = \frac{d - 1}{d} \left(1 - \frac{p}{d} \right)^{d-2}$. The total overlap is the product of local overlaps, $w = \prod_{i=1}^N w_i$. Thus, the moments of $\mathcal{P}(w)$ are directly obtained as the Haar IPR to the power of $N$:
\begin{equation}\label{eq:frame_potential_rps}
I^{(k)}_{\murps} = D \binom{d+k-1}{k}^{-N} \, .
\end{equation}
Notice Eq.~\eqref{eq:frame_potential_rps} coincides with the RMPS result -- Eq.~\eqref{eq:ipr_rnmps_obc} -- in the case 
$\chi=1$ ($r=0$). The distribution of $w$ is the distribution of the product of $N$ i.i.d. random variables.
For qubits ($d=2$), the calculation is easy since each $w_i$ has a flat distribution: $\mathcal{P}(w_i)=\frac{1}{2}$ for $w_i \in [0,2]$. In this case, one can find: 
\begin{equation}
    \mathcal{P}_{\text{IPR}}(w) = \frac{1}{D} \frac{1}{(N-1)!} \left(-\log \left(\frac{w}{D}\right)\right)^{N-1} \, .
\end{equation}
For generic $d$, one can employ a log-normal approximation by noting that $\log w = \sum_{i=1}^N \log w_i$ tends to be Gaussian distributed for large $N$. We have therefore $\mathcal{P}_{\text{IPR}}(w) \simeq \text{Lognormal}(w; N \mu, N \sigma^2)$,
with mean and variance given by
\begin{align}
    \begin{split}
        \mu &= \Ex[\log w_i] = \log d - H_{d-1} \\
        \sigma^2 &=  \Var[\log w_i] = \frac{\pi^2}{6} -\Psi^{(1)}(d) \, .
    \end{split}
\end{align}
Here, $H_i$ is the $i-$th Harmonic number and $\Psi$ is the polygamma function. A plot of the distribution $\mathcal{P}(w)$ for RPS, compared with the PT, is shown in Fig.~\ref{fig:pwrps} for local dimension $d=2,3$ and $N=15$.  

\section{IPR in MPS with OBC}

One can extend the calculation of IPR to the case of Random MPS with periodic boundary conditions (PBC). Although RMPS in PBC are not properly normalised~\cite{lami2024quantum,haferkamp2021emergent}, it is still interesting to write the result, which takes also a form very close to the case of the random phase model. Calculations are performed by computing the transfer matrix in Eq.~\eqref{eq:IPR_bulk}, and its trace to the power $N$. We observe that the matrices $W$ and $G$ commute, so they can be simultaneously diagonalized. In fact, in the basis of projectors onto the representations of $S_k$ we have 
\begin{equation}
    \Wg(d\chi) = \sum_{\lambda\vdash k } c^{-1}_\lambda(d\chi) P_\lambda\;,\qquad G(\chi) = \sum_{\lambda\vdash k } c_\lambda(\chi) P_\lambda\;
\end{equation}
where $\lambda$ are the partitions of the integer, and $c_\lambda(D) = \prod_{(i,j)\in \lambda} (D+i-j)$. They have multiplicity given by the hook formula
\begin{equation}
    f_{\lambda} = \frac{k!}{\prod_{(i,j)\in \lambda} h_{i,j}}\;,
\end{equation}
with $h_{i,j} = (\lambda_i -j)+ (\lambda_j'-i)+1$ and $\lambda_i$ the length of the $i$-th row of the Young tableau associated with $\lambda$, and $\lambda_j'$ the $j$-th column~\footnote{In fact, $f_\lambda = \chi_\lambda(())$ the character of the identity permutation, and $\sum_{\lambda \vdash k} f_\lambda^2 = k!$, consistently.}
It follows that for RMPS with PBC 
\begin{equation}
     \mathcal{I}^{(k)}_{\murmps} = D \sum_{\lambda \vdash k} f_\lambda^2 \left(\frac{c_\lambda(\chi)}{c_\lambda(d\chi)}\right)^N\;.
\end{equation}
However, the distribution $\mathcal{P}(w)$ associated to these moments cannot be written in simple terms. \\

\end{document}